\documentclass[preprint]{aastex}
\usepackage{amsmath}
\usepackage{graphicx}

\newcommand{\fexviii}{[Fe~{\sc xviii}]}
\newcommand{\sixii}{Si~{\sc xii}}
\newcommand{\ovi}{O~{\sc vi}}

\shorttitle{Plasma heating in a post eruption Current Sheet}
\shortauthors{Susino, Bemporad \& Krucker}

\begin{document}

\title{Plasma heating in a post eruption Current Sheet: a case study based on ultraviolet, soft, and hard X-ray data}

\author{Roberto Susino\altaffilmark{1} and Alessandro Bemporad}
\affil{INAF - Osservatorio Astrofisico di Torino, via Osservatorio 20, I-10025 Pino Torinese (TO), Italy}
\email{sur@oact.inaf.it}

\and

\author{S\"am Krucker\altaffilmark{2}}
\affil{University of Applied Sciences and Arts Northwestern Switzerland, CH-5210 Windisch, Switzerland}
\email{krucker@ssl.berkeley.edu}

\altaffiltext{1}{Present address: INAF - Osservatorio Astrofisico di Catania, via Santa Sofia 78, I-95123 Catania, Italy}
\altaffiltext{2}{Also at: Space Sciences Laboratory, University of California, Berkeley, CA 94720-7450, USA}

\begin{abstract}
Off-limb observations of the solar corona after Coronal Mass Ejections (CMEs) often show strong, compact, and persistent UV sources behind the eruption. They are primarily observed by the SOHO/UVCS instrument in the ``hot'' \fexviii\ $\lambda$974~\AA\ line and are usually interpreted as a signature of plasma heating due to magnetic reconnection in the post-CME Current Sheet (CS). Nevertheless, the physical process itself and the altitude of the main energy release are currently not fully understood. 
In this work, we studied the evolution of plasma heating after the CME of July~28, 2004, by comparing UV spectra acquired by UVCS with soft and hard X-ray images of the post-flare loops taken by GOES/SXI and RHESSI. The X-ray data show a long-lasting extended source that is rising upwards, toward the high-temperature source detected by UVCS. UVCS data show the presence of significant non-thermal broadening in the CS (signature of turbulent motions) and a strong density gradient across the CS region. The thermal energy released in the HXR source is on the order of $\sim 10^{32}$ erg, a factor $\sim 2$-5 larger than the energy required to explain the high-temperature plasma sampled by UVCS. Nevertheless, the very different time evolutions of SXR and HXR sources compared to the UV emission suggest that reconnection occurring above the post-eruption arcades are not directly responsible for the high-temperature plasma sampled higher up by UVCS. We conclude that an additional plasma heating mechanism (such as turbulent reconnection) in the CS is likely required.
\end{abstract}


\section{Introduction}

In order to understand the origin and evolution of solar eruptions (namely Coronal Mass Ejections---CMEs) it is of fundamental importance that, on one hand models and numerical simulations based on different hypotheses are performed aimed at reproducing the same observations, and on the other hand analysis of data focuses on the identification of plasma structures envisaged in different models, in order to discriminate between them. 
One of the most important structures that are expected to form in post-CME coronal reconfiguration are Current Sheets (CSs): many CME models predict (regardless of what initiates the eruption) the formation of an elongated CS extending from the top of the post-eruption loop system to the ejected flux rope \citep[e.g,][]{for00,lin04,for06}. 
CSs are regions of large gradient in the magnetic-field tangential component caused by layers of antiparallel field lines where magnetic free energy is converted into thermal and kinetic energies and into acceleration of solar energetic particles \citep[][]{lin00,for06}.
CSs are a necessary part not only of analytic models, but also of numerical simulations \citep[e.g.,][]{ril02} of CMEs, since they allow the CME to escape from the Sun, and create a helical flux rope that might account for magnetic clouds in interplanetary space \citep{gos06}.

Although a CS is supposed to be so thin as to make direct observation quite difficult, in the last decades these features have been identified in white-light data acquired by the Large Angle and Spectrometric Coronagraph \citep[LASCO;][]{bru95} on board the Solar and Heliospheric Observatory (SOHO), and recently also by the STEREO/COR coronagraphs, as long-lived, bright structures extending radially in same direction of the propagating CME \citep[e.g.,][]{cia02,web03,pat11}.
Detection of post-CME CS plasma has been recently reported in EUV and X-rays with Hinode/XRT images \citep[e.g.][]{sav10} and Hinode/EIS spectra \citep[e.g.][]{lan12}, where 3~MK plasma was observed just above the post-flare loops; signatures of CSs plasma were also reported in X-ray data acquired by RHESSI \citep[e.g.][]{sui03}.

Moreover, post-CME CSs have been identified also in UV spectroscopic data acquired in the wake of CMEs by the Ultraviolet Coronagraph Spectrometer \citep[UVCS;][]{koh95} on board the SOHO spacecraft. 
They are typically detected as narrow, very hot (several MK) features, most prominently emitting in the \fexviii\ $\lambda$974~\AA\ line which forms around 5~MK ($\log T\simeq 6.7$). 
Typically this high-temperature emission is observed immediately after the transit of the CME \citep[e.g.,][]{lee06,cia08,sch10} and lasts for many hours or even a few days \citep[e.g.,][]{cia02,ko03,bem06} after the event. 
Recently, \citet{cia13} showed that there is not a one-to-one correspondence between post-CME white-light rays and UV emission: only 40\% of white-light rays (over a sample of 157 rays) showed also UV emission, only 18\% of them were hot features consistent with the CS interpretation; moreover, many CS show UV emission without an associated white-light ray.

Despite these works, the origin of the high-temperature emission observed by UVCS after CMEs is not fully understood. 
Magnetic reconnection occurring in a turbulent plasma has been proposed as a possible solution \citep{bem08}.
Microscopic instabilities inducing plasma turbulence are expected to be able to produce an anomalous resistivity \citep[e.g.,][]{hsu00}, much larger than the classical one, which may account for the high-temperature plasma observed by UVCS. 
Vertically elongated CSs are expected to be unstable mainly via tearing and/or Kelvin-Helmholtz instabilities that may produce magnetic islands via successive reconnections and lead the plasma towards a turbulent state. 
Also stochastic reconnection \citep{laz99} or fractal reconnection \citep[see][]{taj97,shi01} produced by recurring second tearing instability may lead to multiple small-scale local reconnections sites in a turbulent ambient plasma and produce a much higher resistivity. 
Hence, the main idea is that energy is supplied at the largest spatial scales being redistributed via MHD turbulent cascade to smaller scales and finally dissipated. 
It has been showed \citep{bem08} that turbulent reconnection can be able to reproduce at macroscopic level the high CS plasma temperatures, the pressure balance between coronal and CS plasma (needed to explain the stationarity of the CS), and the much larger observed thickness of the CS (broadened by turbulence). 
In agreement with these pictures, significant non-thermal velocities (on the order of 30-60~km~s$^{-1}$), possible signature of plasma turbulence, were detected in many post-CME CSs \citep{bem08,cia08,sch10}.

Alternatively, \citet{ko10} more recently proposed a model \citep[based on the working hypothesis proposed by][]{vrs09} where a steady-state Petschek-like reconnection is assumed to take place in the diffusion region located above the typical post-flare loops. 
In this model the plasma is heated and ejected outwards from the diffusion region, leading to the formation of a collimated ``brush'' of high-temperature plasma expanding in the outer corona; hence, a real current-sheet structure does not form. 
This model is able to reproduce the projected width of the white-light and UV observed features, inferred electron densities, as well as UV and X-ray emissions. 
In agreement with this picture, analysis of the X-ray emission observed by the Geostationary Operational Environmental Satellite (GOES), and RHESSI \citep{lin02}, showed that the X-ray coronal source located at the top of the post-CME reconnected loop system has, at least in one case \citep[see][]{sai09}, sufficient thermal energy to provide the heat for the CS. 
In this scenario, reconnection and plasma heating occur mostly near the loop tops rather than in the CS itself. 

Nevertheless, it is well known that numerical models are, at present, unable to produce a steady-state Petschek reconnection as the one assumed by \citet{ko10}, unless an ``ad hoc'' nonuniform resistivity, enhanced at the X-point of the central diffusion region, is adopted \citep[see][and references therein]{kul01}. 
The origin of this enhanced resistivity is unknown. 
Moreover, although the X-ray coronal source associated with post-flare coronal loops could technically account for the CS plasma heating, it is unclear how reconnection occurring at the base of it could explain at the same time an X-ray emission that fades typically in a few minutes \citep[see, e.g.,][]{pet02} or a few hours \citep{sai09}, and the high-temperature emission observed by UVCS that fades in a few days \citep{ko03,bem06}.

In the present work, we analyze UV and X-ray observations in order to investigate the CS associated with the 2004 July~28 CME.
In this event, a narrow, long-lasting ($\sim 2.5$~days) emission in the \fexviii\ $\lambda$974~\AA\ line was detected by UVCS after the CME passed through its field of view. Concurrent hard and soft X-ray emission was also observed by RHESSI and GOES, probably correlated with the CS but with different temporal characteristics. 
We derive from the UV spectra the electron temperature, density, and the thermal energy content of the plasma embedded in the CS, and compare them with the corresponding quantities inferred from X-ray data. 
We also derive non-thermal line widths of the CS and compare them with expectations based on Petschek and turbulent reconnection pictures \citep[see, e.g.,][]{cia08,bem08}.

\section{The 2004 July~28 Coronal Mass Ejection}

\begin{figure}[h]
\centering\includegraphics[width=\textwidth]{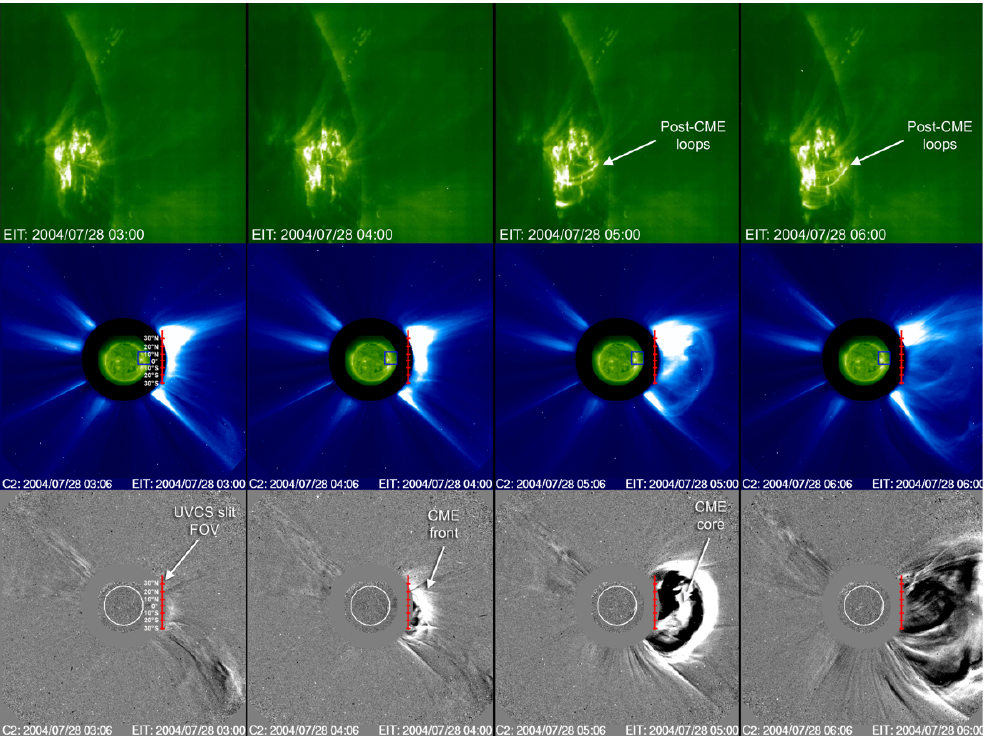}
\caption{Sequence of EIT 195~\AA\ images (top row), composite SOHO/LASCO-C2 and EIT EIT 195~\AA\ images (middle row), and running-difference LASCO-C2 images (bottom row) taken at several times during the July~28, 2004 CME event.
In the top row, EIT images show a close-up on the source active region; post-CME loops are observed to form after $\sim 05$:00~UT. The corresponding region is outlined with a blue box in the middle-row images.
The vertical, red straight line in the middle- and bottom-row images shows the projected location of the UVCS slit Field of View (FOV); left panels give also the latitudes covered along the UVCS slit, for future reference. 
In the bottom row, running differences of LASCO-C2 white-light images are used to enhance the visibility of CME features.\label{event_figure}}
\end{figure}

On 2004 July~28, a slow CME event was detected in white light by the LASCO on board SOHO (Figure \ref{event_figure}). It was classified as a partial halo CME with maximum velocity of $\sim 754$~km~s$^{-1}$ in the online LASCO CME Catalog \citep{gop09}.
The CME originated from active region (AR) NOAA 0652, located at the West limb of the Sun, around 03:00~UT.
At that time, the GOES satellite did not provide evidence of any large flare event, therefore it's difficult to infer the CME starting time.
GOES recorded a gradual increase of the X-ray flux starting at $\sim 02$:30~UT, with peak emission at 05:20~UT. 
Several secondary maxima observed later on in the X-ray light curves were produced by the quite intense flaring activity of the source AR.

LASCO C2 white light (WL) images (see Figure~\ref{event_figure}) show a bright front expanding westwards at a projected latitude of $\sim 14^\circ$NW, followed by a more diffuse and faded structure identifiable as the core of the CME.
The leading edge of the CME was first detected at 03:30~UT, while the core starts to enter in the LASCO/C2 field of view around 04:30 UT.
After the eruption, no bright radial WL feature was observed to connect the CME with the location of the source AR that could be directly identified as the visible signature of a post-CME current sheet; as mentioned in the Introduction, this is not unusual. 
LASCO data also showed repeated transient plasma jets of chromospheric material from the source AR, propagating at a projected latitude of~$\sim 45^\circ$~S.
Due to their location and temporal evolution, however, they seem to be unrelated with the CME and the subsequent formation of the CS.

EUV images acquired in the Fe~{\sc xii} $\lambda$195~\AA\ spectral line with the SOHO Extreme-ultraviolet Imaging Telescope \citep[EIT;][]{del95} show the appearance at 04:36~UT of a post-CME loop system located in the wake of the source AR (see Figure~\ref{event_figure}).
These loops brighten progressively from lower to higher altitudes, reaching a height above the solar surface of $\sim 0.23\ R_\odot \simeq 1.6\times 10^5$~km, as estimated from EIT images, at 14:00~UT.
This is consistent with the standard post-CME picture where magnetic field lines are reconnecting at increasing altitudes after the eruption, leading to the formation of EUV loops at higher altitudes. The loops are quite close to the equatorial region: for instance, around 06:00~UT the tops of the loops cover a projected latitudinal interval between 2$^\circ$S - 5$^\circ$N.
The loop system remains visible for a long time after its formation, at least until 22:00~UT.
After then, the combination of solar rotation and topological changes of the source AR which is continuously evolving makes difficult to clearly identify the loop configuration.
Note that EIT Fe~{\sc xii} images also show several unstructured ejection of hot plasma originating from the southernmost part of the AR and propagating southward: these ejections are the same plasma jets observed later on in WL by LASCO at higher altitudes.

During the second part of the CME/Flare event discussed in this paper, the Nancay Radioheliograph observed radio emissions in the 150-400~MHz range that is likely associated with our event. Solar Geophysical Data classifies this emission as a radio type-I noise storm; however, a more close investigation is required to confirm this.  We did not include radio images in this analysis, but we are currently investigating a possible connection and will potentially publish this findings in a future paper.

\section{UVCS Observations}

\begin{table}[ht]
\begin{center}
\caption{Lines identified in UVCS spectra.\label{lines}}
\begin{tabular}{lllc}
\\
\tableline\tableline
$\lambda_{\rm rest}$ & Ion & Transition & $\log T_{\rm form}$ \\
(\AA)                &     &            & (K)                 \\
\tableline
972.54  & H~{\sc i}      & Ly$\gamma$                              & 4.5 \\
974.86  & \fexviii       & $2s^22p^5~^2P_{3/2}-2s^22p^5~^2P_{1/2}$ & 6.7 \\
977.02  & C~{\sc iii}    & $2s^2~^1S_{0}-2s2p~^1P_{1}$             & 4.8 \\
1025.72 & H~{\sc i}      & Ly$\beta$                               & 4.5 \\
1031.91 & \ovi           & $1s^22s~^2S_{1/2}-1s^22p~^2P_{3/2}$     & 5.5 \\
1037.61 & \ovi           & $1s^22s~^2S_{1/2}-1s^22p~^2P_{1/2}$     & 5.5 \\
520.66  & \sixii         & $1s^22s~^2S_{1/2}-1s^22p~^2P_{1/2}$     & 6.3 \\
\tableline
\end{tabular}
\end{center}
\end{table}

UVCS observations analyzed here started on July~27 at 23:51~UT, and lasted until July~29, 02:03~UT. These data were acquired during a SOHO-RHESSI Joint Observation Program (JOP 153) aimed at detection and study of major flares.
The projected Field of View (FOV) of the UVCS slit has been located perpendicular to the solar equator at a heliocentric distance of 1.77~$R_\odot$ above the west limb (the straight red line in Figure \ref{event_figure}).
The slit width was 100 $\mu$m, and the data were binned over 3~pixels (21$''$ per bin) in the spatial direction and over 2~pixels (0.1986~\AA\ per bin) in the spectral direction.
UVCS recorded spectra with 120~s exposure time, in the spectral ranges \mbox{1022.84-1044.69}~\AA\ and \mbox{972.79-981.53}~\AA.
Table~\ref{lines} lists the lines that have been detected and identified during the observations, together with the formation temperatures of the emitting ions as obtained from the ionization equilibria of \citet{bry06}.
In addition, we also analyzed UVCS synoptic data acquired on both July~29, from 08:29~UT until 10:12~UT, and July~30, from 15:10~UT until 15:29~UT.
In the first synoptic data set, spectra were acquired at five heliocentric distances (1.6, 1.7, 1.9, 2.0, and 3.0~$R_\odot$), while in the second one the slit has been located at three different altitudes (1.6, 1.7, and 2.0~$R_\odot$).
In both cases, the slit central latitude, the slit width, the spatial and spectral resolution, were the same as described above.
Standard calibration of the UVCS data has been performed using the most recent calibration routines included with the Data Analysis Software (DAS), version 4.0.

\begin{figure}[h]
\centering\includegraphics[width=\textwidth]{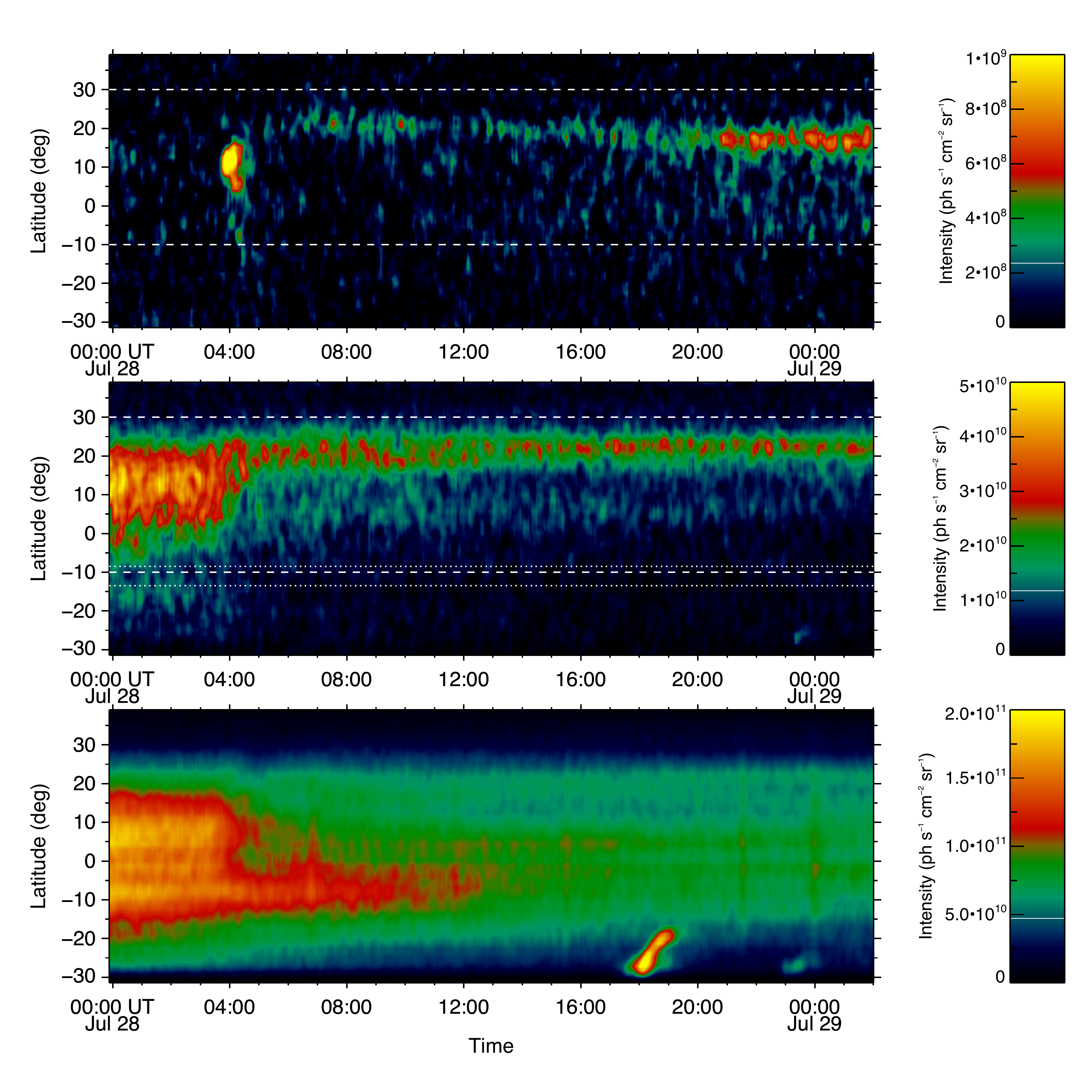}
\caption{Images of \fexviii\ $\lambda974$~\AA, \sixii\ $\lambda520$~\AA, and \ovi\ $\lambda1032$~\AA\ along the slit as function of time. The vertical axis show the position along the slit in latitudes, to be directly compared with latitudes given in Figure~\ref{event_figure}. Horizontal dashed and dotted lines mark the latitudinal interval we used to estimate the average line intensity evolution shown in Figure~\ref{evol} and the average coronal \sixii\ intensity (see text)\label{maps}}
\end{figure}

Every spectral line has been integrated in the spectral direction and the average line background has been subtracted, in order to produce 2D maps of the total line intensity observed at different latitudes along the slit ($y-$axis) and at different times ($x-$axis). Figure \ref{maps} shows the images obtained in the \fexviii\ $\lambda974$~\AA, \sixii\ $\lambda$520~\AA, and \ovi\ $\lambda$1032~\AA lines intensities. Different latitudes along the slit in this Figure can be directly compared with the latitude scale shown in left panels of Figure \ref{event_figure}).
Looking at the \fexviii\ image (showing high temperature plasma), no significant emission is present before $\sim 04$:00~UT of July 28; later on, it is possible to distinguish between two different kinds of coronal emission. 
Nearly at 04:00~UT, an isolated, pronounced bright peak is observed around a latitude of $10^\circ$~N, lasting for about 1~hour.
Shortly after ($\sim 05$:00~UT), an increasingly bright feature appeared $\sim 10^\circ$ northward with respect to the initial peak.
Over about 20~hours, the bulk of this \fexviii\ ``tail'' emission moved southwards by $\sim 5^\circ$. 

Because of the very high maximum formation temperature of the Fe$^{17+}$ ion (around $10^{6.7}-10^{6.8}$~K, depending on the ionization equilibrium assumed), the \fexviii\ $\lambda974$~\AA spectral line is usually not observed in the intermediate corona. Hence, as mentioned in the Introduction, in all previous works dealing with UVCS data this high temperature emission observed after the transit of CMEs has been interpreted as a signature of plasma heating due to magnetic reconnection occurring in the post-CME Current Sheets \citep[][]{lee06,cia08,sch10,cia02,ko03,bem06}. For this reason, in what follows we will simply call ``CS region'' the latitudinal interval where the \fexviii\ ``tail'' emission is observed; a more detailed identification of the real CS location in the corona will be discussed in more details later.

The \sixii\ image (indicative of plasma around $10^{6.3}$~K) shows a diffuse, bright emission localized between the equatorial plane and a latitude of $20^\circ$~N, that precedes the \fexviii\ peak emission and is probably produced by the pre-CME coronal streamer crossing the UVCS FOV at that latitudes (see Fig.~\ref{event_figure}).
After $\sim 04$:00~UT, substantial reduction of \sixii\ emission occurs and new fainter emission is detected in two narrow regions located at the latitudes of $\sim 5^\circ$~S and $\sim 20^\circ$~N, i.e. almost symmetrically around the position of the initial \fexviii\ peak ($10^\circ$~N).
The intensity of the \sixii\ line in the northern region is about 4~times larger than that in the southern one, while in the central part it is significantly dimmed.
The ``tail'' CS emission observed in the \fexviii\ line partially overlaps with the dimmed region observed in the \sixii\ line, at least in the last 2~hours of observations.

The third panel of Figure~\ref{maps} shows the evolution of \ovi\ emission, representative of cooler plasma.
Before 04:00~UT, a very diffuse emission partially cospatial with the \sixii\ emission is observed which is associated with the pre-CME coronal streamer.
After the appearance of the \fexviii\ peak, the emission in the \ovi\ line becomes fainter and is located mainly at lower latitudes with respect to the \fexviii\ ``tail'' CS emission, but just until 12:00~UT.
At 18:00~UT and 23:00~UT two bright knots of emission appear at lower latitudes ($30^\circ$~S-$20^\circ$~S), likely associated with two of the plasma jets produced by the flaring active region and observed by LASCO-C2 in WL. 
Hence, for this event there is no clear UV signature of the transit of the CME front and core, but only a decrease of the \ovi\ and \sixii\ emissions likely associated with a coronal density decrease due to the CME expansion.

\begin{figure}[h]
\centering\includegraphics[width=0.75\columnwidth]{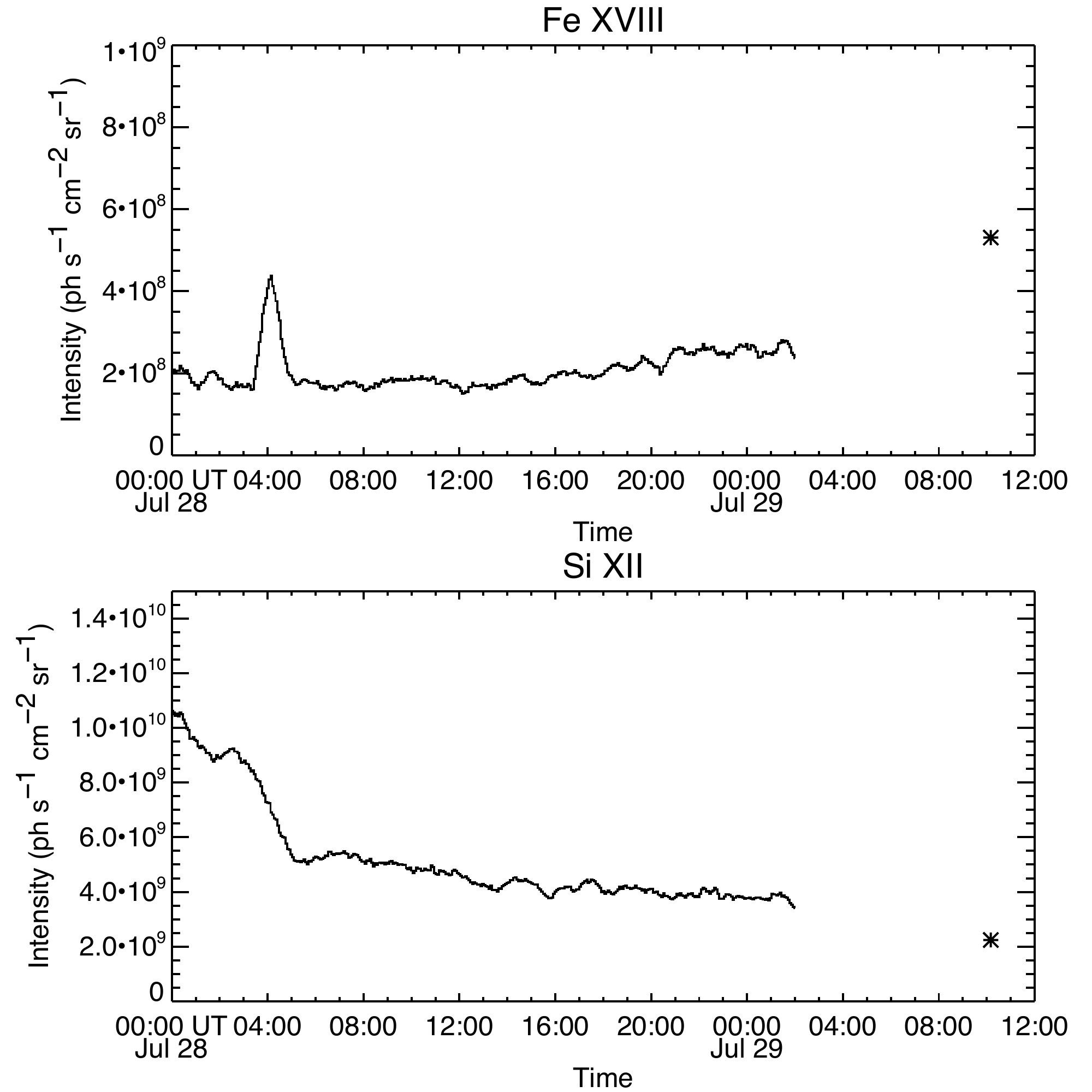}
\caption{\fexviii\ and \sixii\ total intensities as a function of time in the latitude interval identified as the post-CME Current Sheet region. The intensities are obtained for a $40^\circ$ wide portion along the slit centered at a latitude of $10^\circ$N, i.e. in the latitudinal region marked by the dashed white lines in Figure~\ref{maps}. The asterisk marks the intensities measured from the synoptic data acquired on July~29.\label{evol}}
\end{figure}

In Figure~\ref{evol} we report the total intensities of the \fexviii\ and of the \sixii\ lines in the CS region, as a function of time.
These intensities are computed in a 40$^\circ$ wide band, centered around $10^\circ$~N, in order to include both the peak and the ``tail'' emission detected in the \fexviii\ line, as well as the emission features observed in the \sixii\ line.
We also reported in the plot the intensity measurement obtained from the synoptic observation performed on July~29. These plots show a clear trend:
after 10:00~UT of July 28, the \fexviii\ line intensity increased continuously during the rest of the observations by about a factor of 3.
Conversely, the \sixii\ line intensity decreased by $\sim 50\%$ in the same time interval.
Note that this behaviour is different from what reported by previous observations of current sheets with UVCS \citep[e.g.,][]{bem06,cia08}, which show that the increase in the \fexviii\ line intensity is generally followed by a decrease, as the CS progressiverly disappears.
In our case, we lack continuos observations between the July~29 synoptic and the successive one performed on July~30, in which, on the other hand, no signature of \fexviii\ emission has been found.
Hence, we can only state that the decreasing phase should have occurred between the times of the two synoptic observations, hence between July~29, 08:29-10:12~UT, and July~30, 15:10-15:29~UT.

As we better explain in the next section, the opposite trend of the intensity curves of the two ions can be explained mainly by a temporal evolution of the plasma temperature in the coronal region sampled during the observations.

\section{Electron temperature and density diagnostics}

The changes in the line emissions detected by UVCS are in general the consequence of the evolution of the plasma electron temperature and density: possible intensity changes due to variation of the elemental abundances are usually neglected.
As already done by previous authors \citep[see, e.g.,][]{cia08}, we use the \fexviii\ to \sixii\ line ratio to estimate the electron temperature $T_e$ of the CS region.
Both spectral lines are due to collisional excitation, followed by spontaneous emission, whose intensity is given by:
\begin{equation}\label{int1}
I_{\rm line}=\frac{1}{4\pi}\int_{\rm LOS}\epsilon_{\rm line}(T_e,A_X)\,n_e n_{\rm H}\,dz\ \text{(phot. s$^{-1}$ cm$^{-2}$ sr$^{-1}$)},
\end{equation}
where $\epsilon_{\rm line}(T_e, A_X)$ (phot cm$^{3}$s$^{-1}$) is the line emissivity which depends mostly on the electron temperature $T_e$ (K) through the ionic fraction and the electron excitation rate, and weakly on the elemental abundance $A_X$, $n_e$ (cm$^{-3}$) is the electron density, $n_{\rm H}$ (cm$^{-3}$) the proton density, and the integration is performed along the line of sight (LOS) through the optically thin coronal plasma.
Restricting to the CS region, where the plasma can be assumed to be isothermal along the LOS, and taking $n_e\approx n_{\rm H}$, Eq.~(\ref{int1}) is usually rewritten as:
\begin{equation}\label{int2}
I_{\rm line}\simeq\frac{1}{4\pi}\epsilon_{\rm line}(T_e, A_X)\,n_e^2\,D,
\end{equation}
where $D$ (cm) is the thickness of the CS along the LOS. Because the quantity $D$ is unknown, in this work we assumed that the extension along the LOS of the CS region is equal to the angular width of the CS along the UVCS slit ($\approx 5^\circ$). In any case, since $n_e \propto 1/sqrt{D}$, even a one order of magnitude uncertainty on $D$ will give an indetermination by a factor $\sim 3$ on $n_e$.
The electron temperature can be then computed by comparing the measured ratio $I_{\rm line1}/I_{\rm line2}$ between the observed line intensities with the expected ratio between the corresponding line emissivities $\epsilon_{\rm line1}(T_e, A_{X1}) / \epsilon_{\rm line2}(T_e, A_{X2})$. 
Equation~(\ref{int2}) can be used in turn to derive also the electron density $n_e$, given the electron temperature $T_e$ derived with the line ratio technique, the observed intensity of one of the two lines, $I_{\rm line}$, and the LOS thickness, $D$. Note that Eq.~(\ref{int2}) assumes a filling factor of unity, hence the resulting densities have to be considered as a lower limit estimate.

\begin{figure}[h]
\centering\includegraphics[width=0.75\columnwidth]{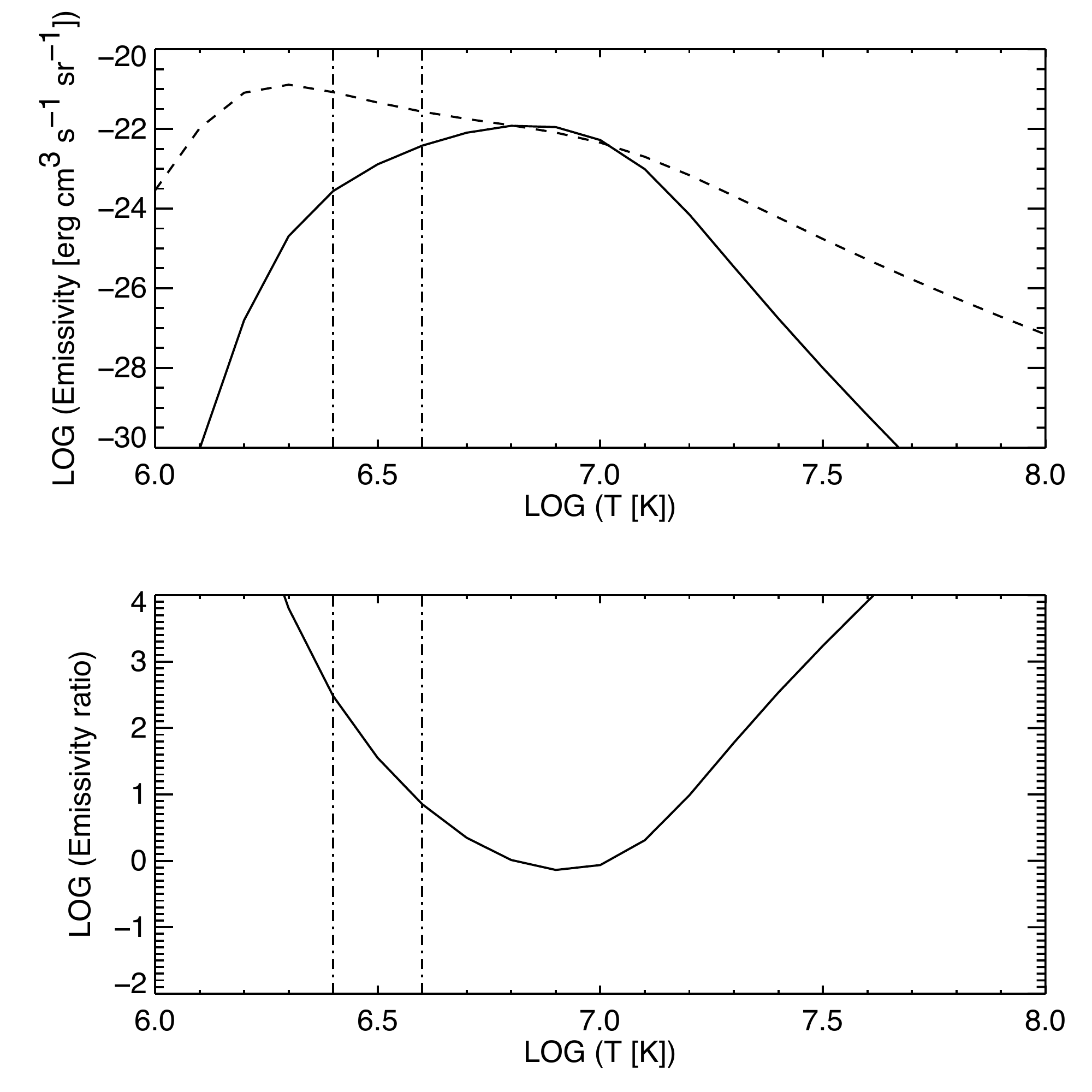}
\caption{Top panel: emissivities of the \fexviii\ $\lambda974$~\AA\ (solid) and \sixii\ $\lambda520$~\AA\ (dashed) lines from the CHIANTI spectral code (ver. 7.1), based on the ionization equilibria of \citet{bry06}. Bottom panel: ratio of the \sixii\ to the \fexviii\ line emissivity. In both panels, the dot-dashed vertical lines mark the temperature range of interest in the present analysis.\label{gfunctions}}
\end{figure}

In our calculations, the line emissivities for the considered transitions are taken from CHIANTI version 7.1 with the ionization equilibria of \citet{bry06} and coronal abundances from \citet{ray97}. Corresponding emissivities are shown in Figure~\ref{gfunctions}. Hence, as also done by previous authors, in this work we assumed ionization equilibrium: implications on the final results of possible deviations from ionization equilibrium are discussed in the last Section.
Furthermore, by using Eq.~(\ref{int2}) we are assuming that the emission is coming from a uniform and isothermal volume, hence derived physical parameters have to be considered as representative of the average CS plasma conditions. 
However, the limited set of lines detected by UVCS in the considered observations (see Table~\ref{lines}) does not allow for a more involved temperature analysis, e.g., using the differential emission measure.

The derivation of the electron temperature is not straightforward, because emission from the quiet, low-temperature corona surrounding the CS region is superposed along the LOS onto the high-temperature emission coming from the CS region and detected in the \fexviii\ and \sixii\ lines.
We make the simplifying assumption that the \fexviii\ is emitted only from the high-temperature plasma inside the CS.
In fact, due to its high-temperature formation ($10^{6.8}$~K), the \fexviii\ line is unlikely to form in quiet corona surrounding the CS, whose mean temperature is expected to be of the order of $\sim 10^6$~K.
Therefore, the low-temperature contribution from the quiet corona is significant only for the \sixii\ line, which has a maximum of line emissivity around $\log T \simeq 6.3$ and a very broad temperature distribution (Figure~\ref{gfunctions}), making this line quite sensitive also to 1~MK plasma.

The major source of uncertainty in the determination of the electron temperature arises from the fact that the ranges of maximum \sixii\ and \fexviii\ line emissivity are only partially overlapping in temperature (see Figure~\ref{gfunctions}).
This means that the signal in one of the two lines will be very small, when there is plasma within the range of significant response of the other line.
For instance, in the range of electron temperature between $10^{6.4}$~K and $10^{6.6}$~K, the ratio of the \sixii\ to the \fexviii\ line emissivity can reach the value of $\sim 10^2$ (see bottom panel of Figure~\ref{gfunctions}), hence in order to increase the signal-to-noise ratio, it is necessary to average the observed intensities over many exposures and spatial bins.

For this reason, we averaged the spectra over temporal intervals of one hour and spatial elements of 2~bins ($42''$). The quiet corona contribution to the \sixii\ line was estimated for each time interval by integrating the \sixii\ spectral profile in a region $5^\circ$ wide located around $11^\circ$~S, i.e., in a portion of the UVCS slit where the projected FOV is located at the same altitude in the corona of the high-temperature \fexviii\ emission (see Figure \ref{maps}). At this latitude the corona shows no significant evolution and we can make the simplifying assumption that the emission detected at this latitude and this altitude in the quiet corona (region marked by dotted lines in Figure~\ref{maps}) is comparable to the unknown coronal emission from plasma at the same altitude, but at the latitude of the CS region. We then subtracted the \sixii\ intensity of the quiet corona to the total intensity measured at the location of the CS region. 

\subsection{CS identification}

\begin{figure}[h]
\centering\includegraphics[width=\textwidth]{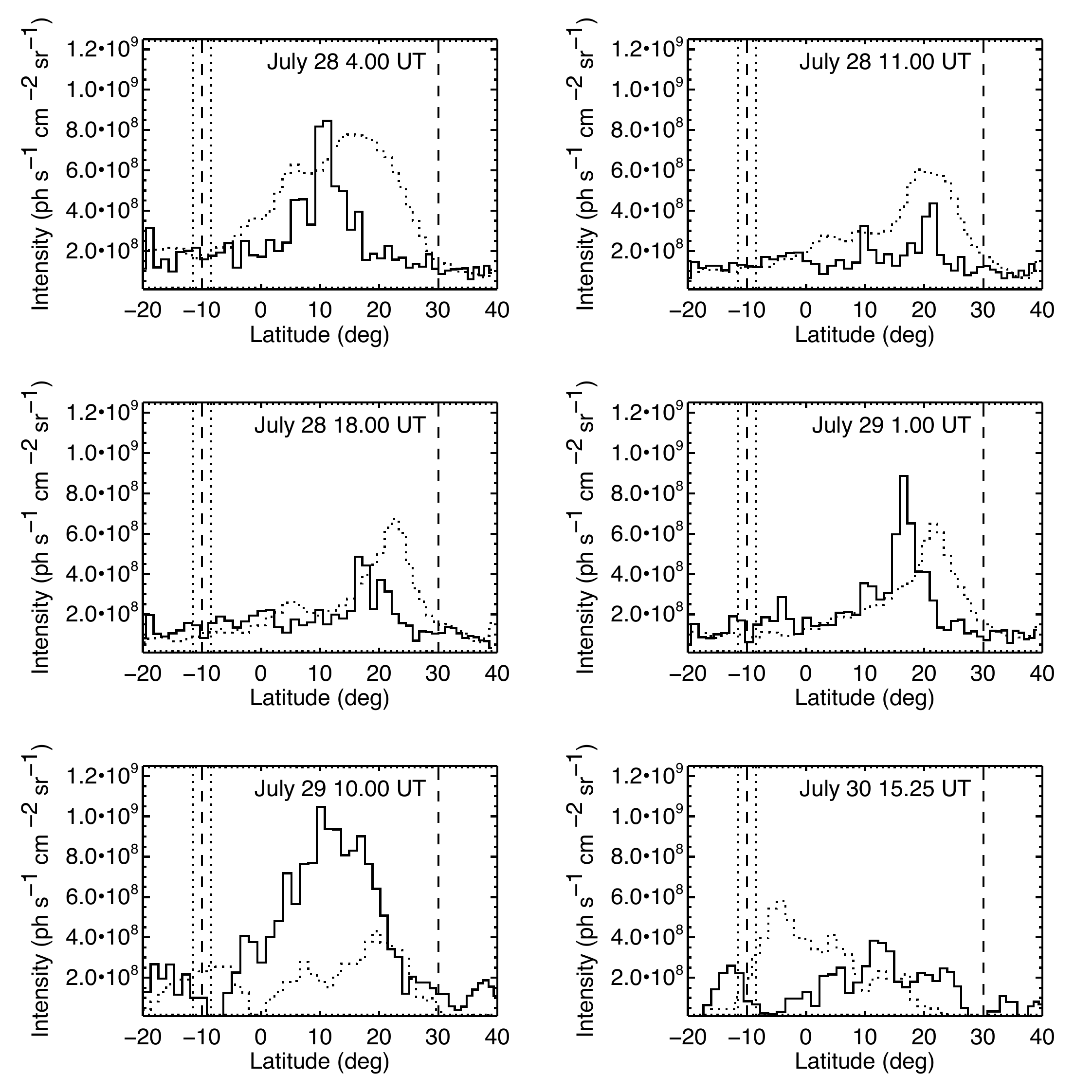}
\caption{\fexviii\ (solid) and \sixii\ (dotted) line intensity distribution, as a function of the position along the UVCS slit at different times during the observations. Intensities have been integrated over 2~hour time intervals. Bottom panels refer to the synoptic observations of July~29 (left) and July~30 (right), both made at a heliocentric distance of $1.7~R_\odot$. The \sixii\ intensity has been divided by a factor of 50 in order to be shown on the same plot.\label{int_dist}}
\end{figure}

The latitudinal distributions of the \fexviii\ and \sixii\ line intensities along the UVCS slit at different times are shown in Figure~\ref{int_dist}.
At 04:00~UT (top left panel), the \fexviii\ line emission is concentrated in a latitude interval $\sim 7^\circ$ wide centered around $11^\circ$~N, while the \sixii\ line emission is more widespread along the slit.
Subsequently, the \fexviii\ line shows a weak emitting feature localized at $\sim 20^\circ$~N whose intensity progressively rises with time.
Eventually, at 10:00~UT on July~29, an enhanced emission in the line is concentrated in a wider latitude interval around $\sim 12^\circ$~N, nearly at the same position of the initial peak at 04:00~UT on July~28.
This can be considered as the last signature of the CS, because during the synoptic observation of July~30 (bottom right panel), a well-defined \fexviii\ intensity pattern is hardly identifiable, even though a local maximum is still present at $\sim 12^\circ$~N.
As for the \sixii\ line, after 04:00~UT on July~28 it exhibits an intensity maximum located at $\sim 22^\circ$~N that remains quite constant until 01:00~UT on July~29; then the line intensity distribution along the slit becomes very irregular, without  features that could be significantly associated with the CS (bottom panels).

The plots show interestingly that the main intensity peaks in the two lines are not coincident, but are shifted by about $7^\circ$.
This is particularly visible at 18:00~UT on July~28 and 01:00~UT on July~29 (middle panels).
This result, which is similar to that found, for instance, by \citet{ko03}, suggests that the CS position cannot be directly associated with the location of the maximum of the \fexviii\ emission, as is usually done, because the asymmetrical distribution of the intensities is most probably caused by a perpendicular density gradient across the CS region, with a consequent lack of correlation between the location of the maximum plasma temperature and that of the maximum of the \fexviii\ emission.

Recently, \citet{mur12} have studied 2D MHD models of line-tied asymmetric current sheets where the reconnection process is made asymmetric in the direction perpendicular to the CS axis by allowing the initial magnetic field strengths and densities to differ.
These models show a significant contrast in plasma pressure and density between the two sides of the CS. 
In particular, the plasma pressure buildup is almost entirely contained on the weak-field side of the CS, and this can justify the asymmetry in the emission line intensities that are expected from the CS region.

\begin{figure}[h]
\centering\includegraphics[width=\textwidth]{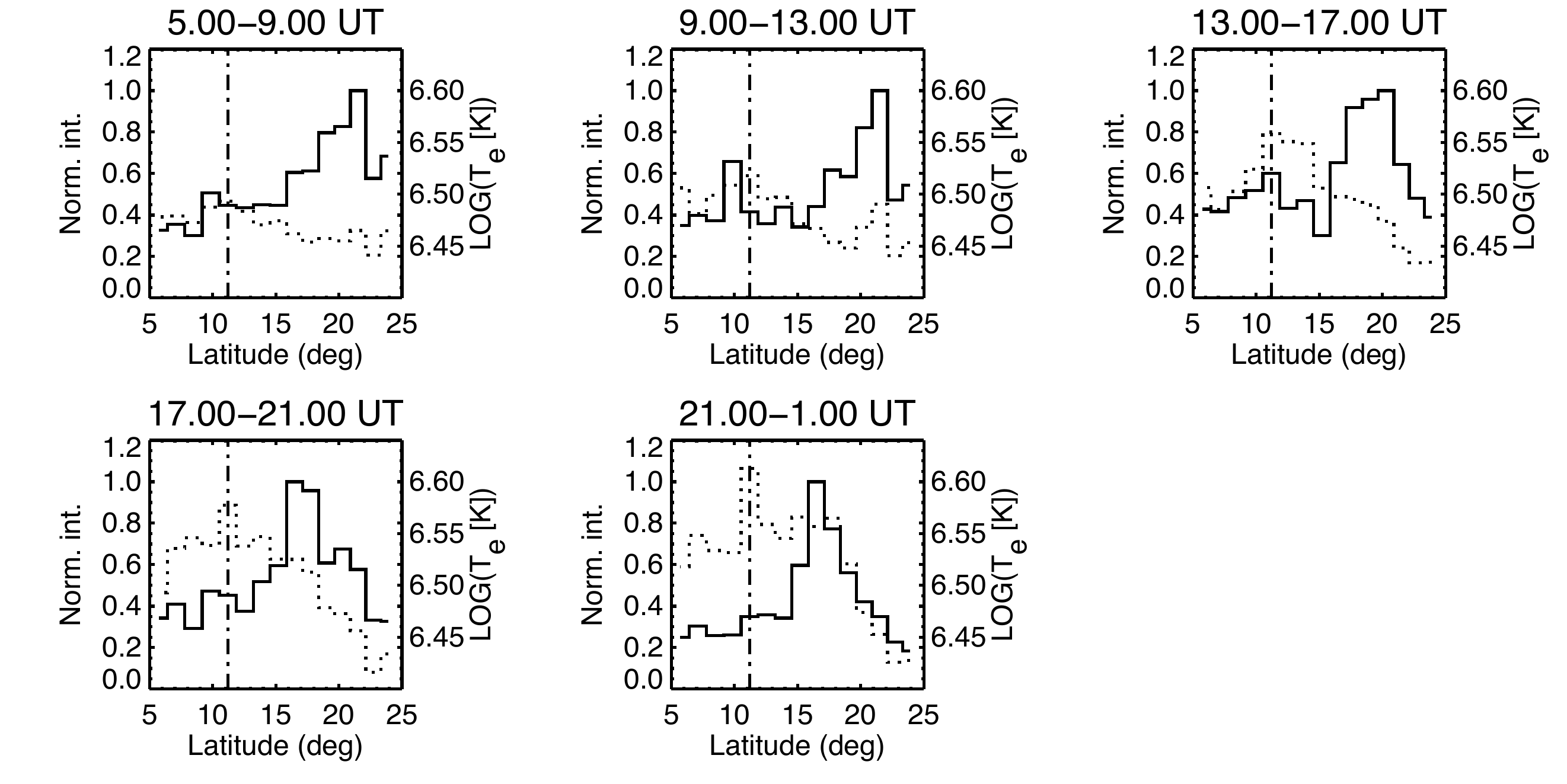}
\caption{\fexviii line intensity distribution (solid) and electron temperature distribution (dashed) derived from the ratio of \fexviii\ and \sixii\ line intensities, as functions of the position along the UVCS slit at different times. The intensities used to perform the calculations have been integrated over time interval of 4~hours. The vertical line indicates the position of the maximum electron temperature.\label{lat_te}}
\end{figure}

In the light of this work, we analyzed the latitudinal distribution of the electron temperature and density in the CS region, at different times, in order to correctly identify the location of the high temperature region, hence the real location of CS region.
The electron temperatures at different latitudes are deduced using the intensity ratio technique described above.
The corresponding results are plotted in Figure~\ref{lat_te}: the temperature peak turns out to be shifted by about $10^\circ$ southward ($\sim 2 \times 10^5$~km) with respect to the maximum of the \fexviii\ intensity. Hence, the center of the high temperature region is much closer to the equator than what simply shown by the \fexviii\ line intensity distribution: this location is in better agreement with the almost equatorial location of the post-CME loops seen in EIT images.
Figure~\ref{lat_te} also shows that an average temperature of $\sim 10^{6.55}$~K is inferred in the centre of this high temperature region.
This value is in-between the formation temperatures of the \sixii\ line ($\log T \simeq 6.3$) and the \fexviii\ line ($\log T \simeq 6.7$) and is only slightly lower than previous estimates of CS temperatures at comparable heights.
For instance, \citet{cia08} reported $T_e$ ranging in $6.60 \leq \log T \leq 6.90$ for the CS they analyzed. 

The latitudinal region where the $T_e$ peak is located corresponds to the real location of the post-CME CS: hence, hereafter in the rest of this work, we use the term ``CS'' to refer to this region. The position of the CS does not vary very much as time progresses and is located around $11^\circ$~N.
A comparison with Figure~\ref{maps} shows interestingly that this is also the average position of the \sixii\ dimmed emission region. This suggests that
the two peaks detected in the \sixii\ line emission are a consequence of the \sixii\ emissivity changing along the slit as temperature decreases northward and southward of the latitude of the maximum temperature.
In fact, as the plasma temperature decreases at higher and lower latitudes, the emissivity of the \sixii\ line progressively increases, since the CS temperature ($\sim 10^{6.55}$~K) is greater than the \sixii\ peak temperature ($\sim 10^{6.3}$~K), then it decreases when the plasma temperature goes below $10^{6.3}$~K.
Conversely, the emissivity of the \fexviii\ line monotonically decreases, because the CS temperature is already lower than the \fexviii\ peak temperature ($\sim 10^{6.8}$~K) (see Figure~\ref{gfunctions}).
Therefore, the increase of the \sixii\ emissivity occurring on both sides of the CS, together with the density gradient occurring in the north-south direction, produce the two intensity peaks observed in \sixii\ image.

\begin{figure}[h]
\centering\includegraphics[width=\textwidth]{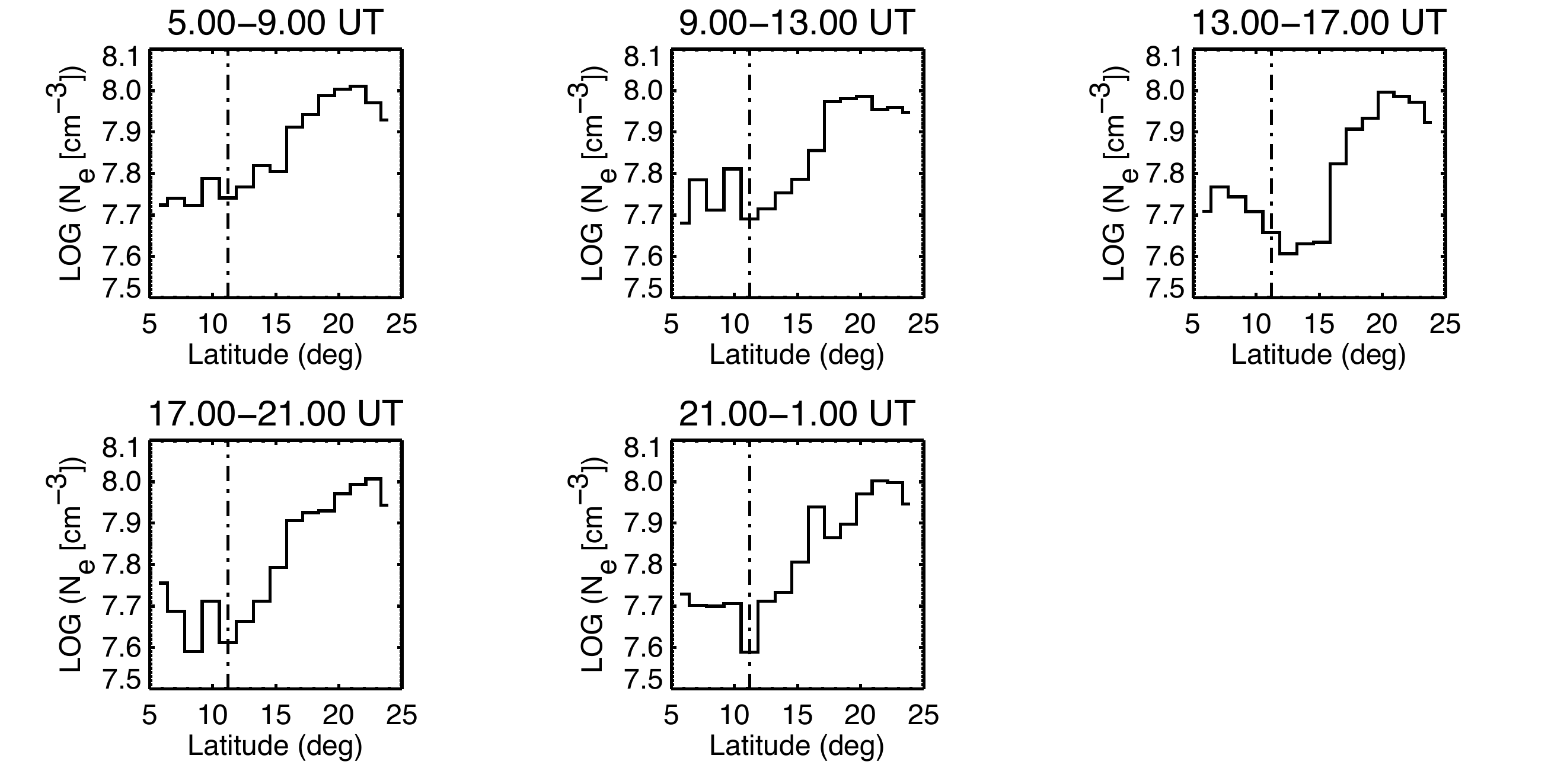}
\caption{Electron density distribution as function of latitude along the UVCS slit, derived from the temperature profile of Figure~\ref{lat_te} in the same time intervals.\label{lat_ne}}
\end{figure}

Given the \sixii\ line intensity emitted at the CS region and the temperature profiles of Figure~\ref{lat_te}, the CS electron densities have been estimated: resulting densities are shown in Figure~\ref{lat_ne} for the same time intervals.
The CS region is characterized by the presence of a density gradient in the direction perpendicular to the CS axis which reflects the asymmetrical distribution of the coronal emission observed in the \fexviii\ and \sixii\ lines.
The variation of the electron density across the CS is about a factor~2.
Such a density gradient is very similar to that found by \citet{mur12} in their simulations of asymmetrical sheets (see Figure~4 of their paper).
Note that the highest temperature in the CS is coincident with an electron density minimum.
This result is different from those found by previous UVCS studies.
In fact, CS plasma is generally found to be denser than the surrounding ambient plasma \citep[see, e.g.,][]{cia08}.
On the other hand, \citet{lan12} recently reported results from the analysis of a post-CME coronal region with Hinode data in which the CS electron density is found to be lower than in the background plasma by a factor of about 1.5-2.

\subsection{Pressure balance analysis}

The latitudinal density and temperature distribution results in a thermal pressure imbalance between the regions located around the high temperature region.
Hence, given the stationarity of the feature reported here, we assumed that the total pressure (thermal plus magnetic) is conserved between the two sides of the CS region, i.e.:
\begin{equation}
	p_{\rm tot}=p_{\rm th}+p_{\rm mag}=2n_ek_{\rm B}T_e+B^2/2\mu=const.,
\end{equation}
where $k_{\rm B}$ (erg K$^{-1}$) is the Boltzmann constant, $B$ (G) the magnetic field strength, and $\mu=4\pi$ the plasma magnetic permeability in cgs units.
We derived a unique thermal pressure profile using the electron temperature and density distributions obtained averaging all the exposures since 05:00~UT to 01:00~UT of July~29 and fitting the resulting curve with an ``ad hoc'' function.
We then calculated a mean value for coronal magnetic field, at the altitude of the UVCS slit, using the empirical formula reported by \citet{dul78}:
\begin{equation}
	B(h)=\frac{1}{2}\cdot(h-1)^{-3/2},
\end{equation}
where $h$ is the heliocentric distance in solar radii and $B$ is expressed in Gauss units.
We inferred the constant value for the total pressure on one side of the CS by summing the magnetic pressure obtained using the mean magnetic field strength and the average value of the thermal pressure obtained from the data.

\begin{figure}[h]
\centering\includegraphics[width=0.75\columnwidth]{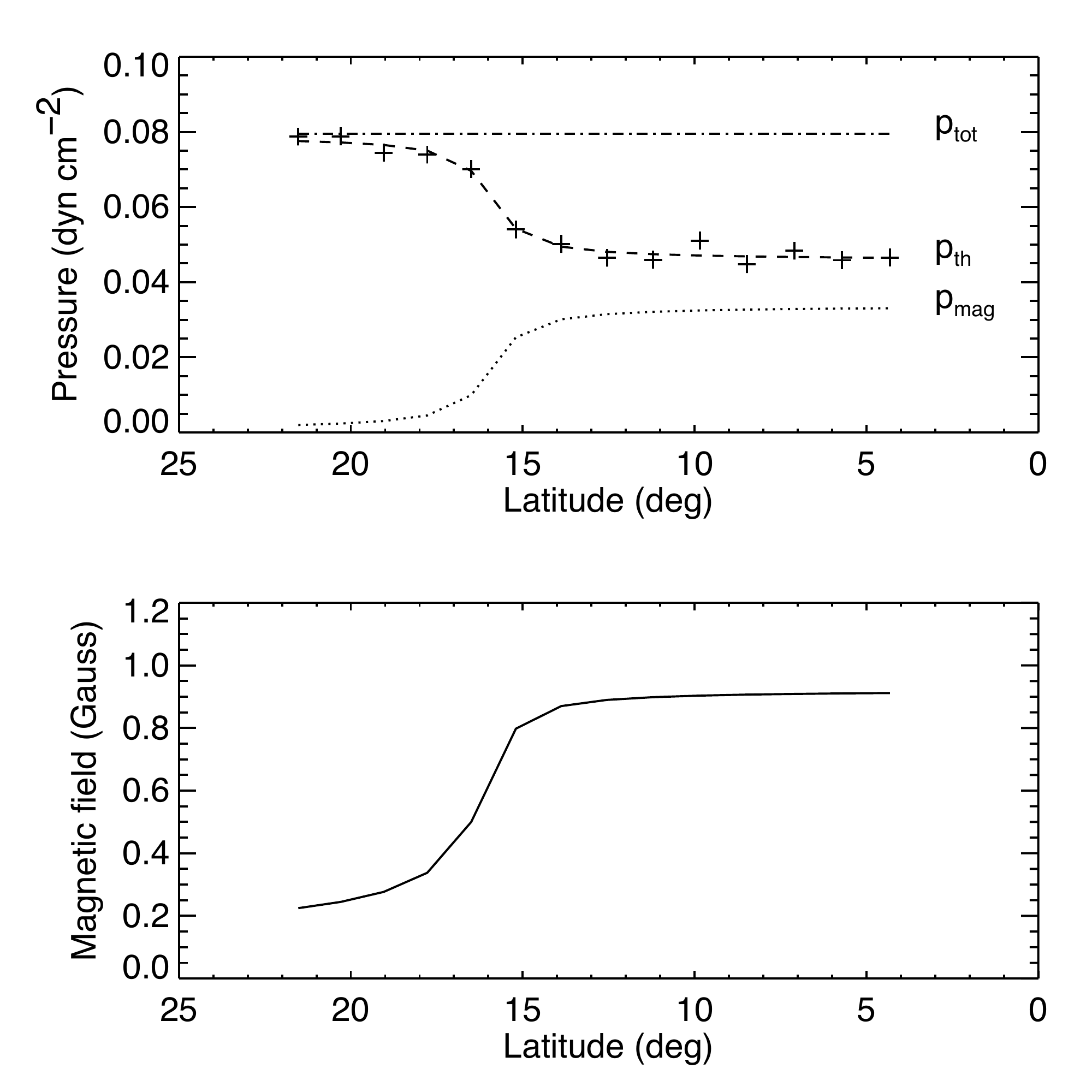}
\caption{Top panel: total plasma pressure (dot-dashed line), thermal plasma pressure distribution derived from the electron temperature and density distributions obtained using the \fexviii\ and \sixii\ line intensity ratio (crosses), fit to the empirical data (dashed line), and magnetic plasma pressure (dotted line). Bottom panel: strength of the magnetic field component parallel to the CS axis obtained from the distribution of the magnetic plasma pressure. All quantities are plotted vs. the position along the UVCS slit.\label{lat_b}}
\end{figure}

Figure~\ref{lat_b} reports the resulting distributions of the thermal pressure, the magnetic pressure, and the total pressure (top panel), and the derived latitudinal variations of the magnetic field strength across the CS (bottom panel).
The total magnetic field latitudinal variation at the two sides of the CS is a factor $\sim 3.6$.
Note that, since the magnetic flux also conserves across the CS \citep[see Eq.~(26) of][]{mur12}, the ratio of the magnetic strength between the two sides of the CS turns out to be inversely proportional to the ratio of the corresponding plasma inflow velocity components.
Note also that the resulting plasma-beta parameter, defined as the ratio of the thermal pressure to the magnetic pressure ($\beta = p_{\rm th}/p_{\rm mag}$), is $\beta \simeq 20-25$ at the northward side and $\beta \simeq 1.4$ at the southward side of the CS.

\subsection{Plasma thermodynamics}

\begin{figure}[h]
	\centering\includegraphics[width=0.75\columnwidth]{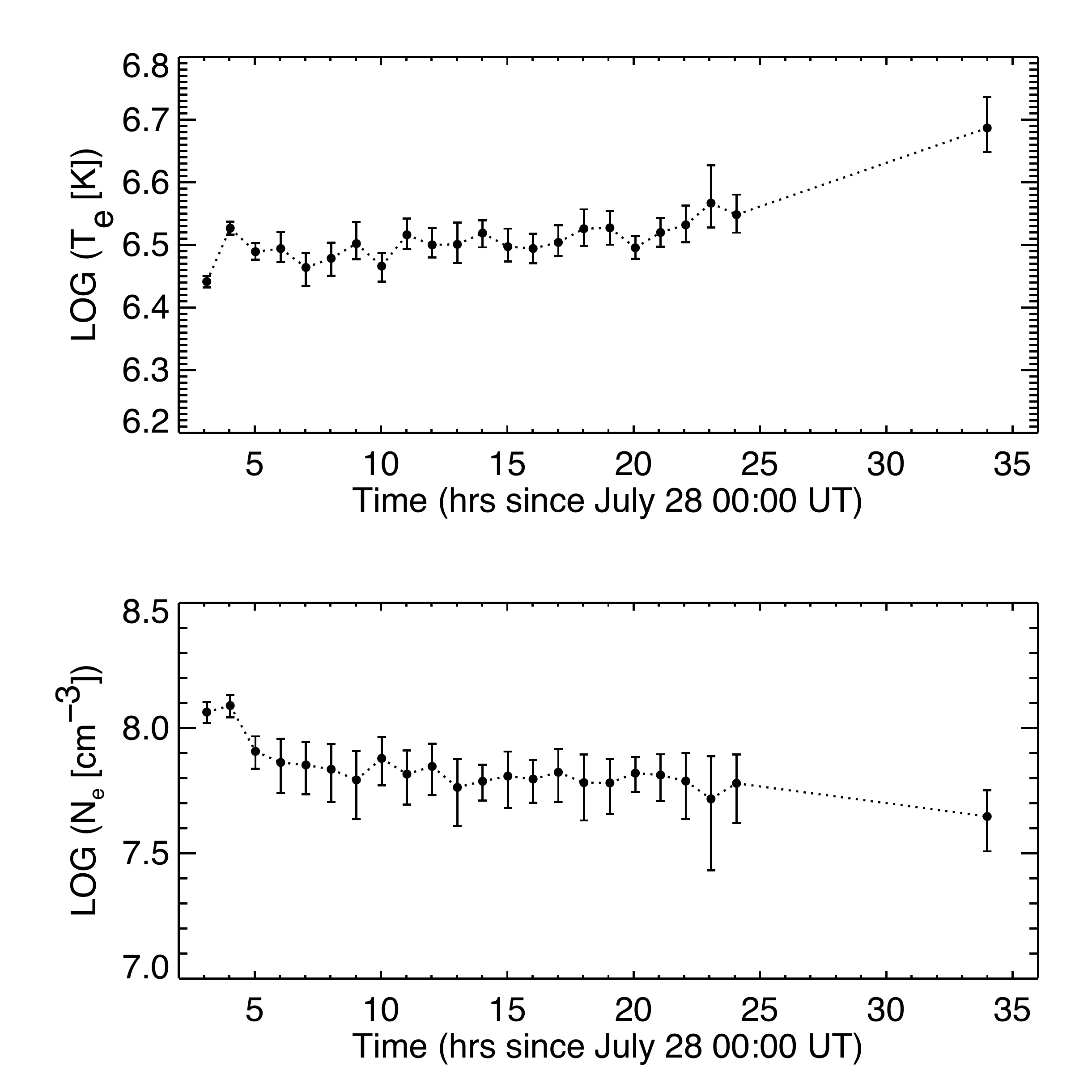}
	\caption{Evolution of the electron temperature (left panel) and density (right panel) as function of the time.\label{uvcs_thermo}}
\end{figure}

Figure~\ref{uvcs_thermo} presents the temporal evolution of the average CS plasma temperature and density: data have been accumulated over a $5^\circ$ wide latitude interval centered at a latitude of $11^\circ$~N, hence over the real high temperature CS region.
The uncertainties in the electron temperatures ($\Delta T_e$) were estimated directly from the uncertainties in the measured line intensities $\Delta I_{\rm Si}$ and $\Delta I_{\rm Fe}$ (derived through Poissonian statistics and standard uncertainty propagation rules) using the relationship
\begin{equation}
(T_e \pm \Delta T_e)= R^{-1}\left(\frac{I_{\rm Si} \pm \Delta I_{\rm Si}}{I_{\rm Fe} \mp \Delta I_{\rm Fe}}\right),
\end{equation}
where $R$ is the theoretical ratio between the \sixii\ and \fexviii\ line emissivities. Resulting $\Delta T_e$ are plotted as error bars in the Figure ~\ref{uvcs_thermo}. The uncertainties in the electron density and in the other thermodynamic quantities were then computed using the standard error propagation formulae.

A $\sim 50\%$ increase in the electron temperature of the CS occurs over the 30 hours of observations (including the measurement from the synoptic observation of July~29).
At the same time, the density inside the CS slowly decreases: initially, until $\sim 08$:00~UT, very rapidly, then more gradually.
The electron density values span in the $7.7 \le \log n_e \le 8.1$ range and are comparable with those measured, for instance, by \citet{bem06} and \citet{cia08} from UVCS data acquired at the same heliocentric distances.
However, the uncertainties in the density determination are very high and the decreasing trend may not be significant.

We used these plasma parameters to estimate the rate of thermal energy flowing in the CS from lower layers at the altitude of the UVCS field of view.
This can be done by multiplying the thermal energy density of the CS plasma, $E_{\rm th}=3n_e k_{\rm B} T_e$, by the volume spanned by the plasma flowing along the CS and across the UVCS field of view in a time unit, $V=(\pi D^2) \cdot v_{\rm out}$. 
Here, $D$ is the half thickness of the CS on the plane of the sky and $v_{\rm out}$ is the outflow velocity of the plasma in the radial direction.

We assumed $D=10^5$~km (corresponding to an angular width of $5^\circ$).
In order to estimate the plasma outflow velocity, the Doppler-dimming analysis of the \ovi\ doublet line intensity can be in principle used.
The ratio between the $\lambda 1032$~\AA\ and the $\lambda 1037$~\AA\ \ovi\ line intensities is an indicator of plasma motions in the radial direction.
However, the outflows in the CS cannot be directly determined from the Doppler dimming of oxygen lines, because the \ovi\ emission originates almost entirely in the quiet, low-temperature corona ($\log T\simeq 5.5$).
Therefore, we crudely estimated the plasma outflow velocity in the surrounding corona using the pre-CME coronal spectra: it turned out $v_{\rm out} \approx 54$~km~s$^{-1}$.
This value is quite low compared to typical stream velocities deduced in the corona, for instance, above the cusp of coronal streamers ($\sim \text{100-150}$~km~s$^{-1}$).
Flows in the CS are more probably faster than in the quiet corona, so our estimate of CS thermal energy is only a lower-limit estimate.

\begin{figure}[h]
	\centering\includegraphics[width=0.75\columnwidth]{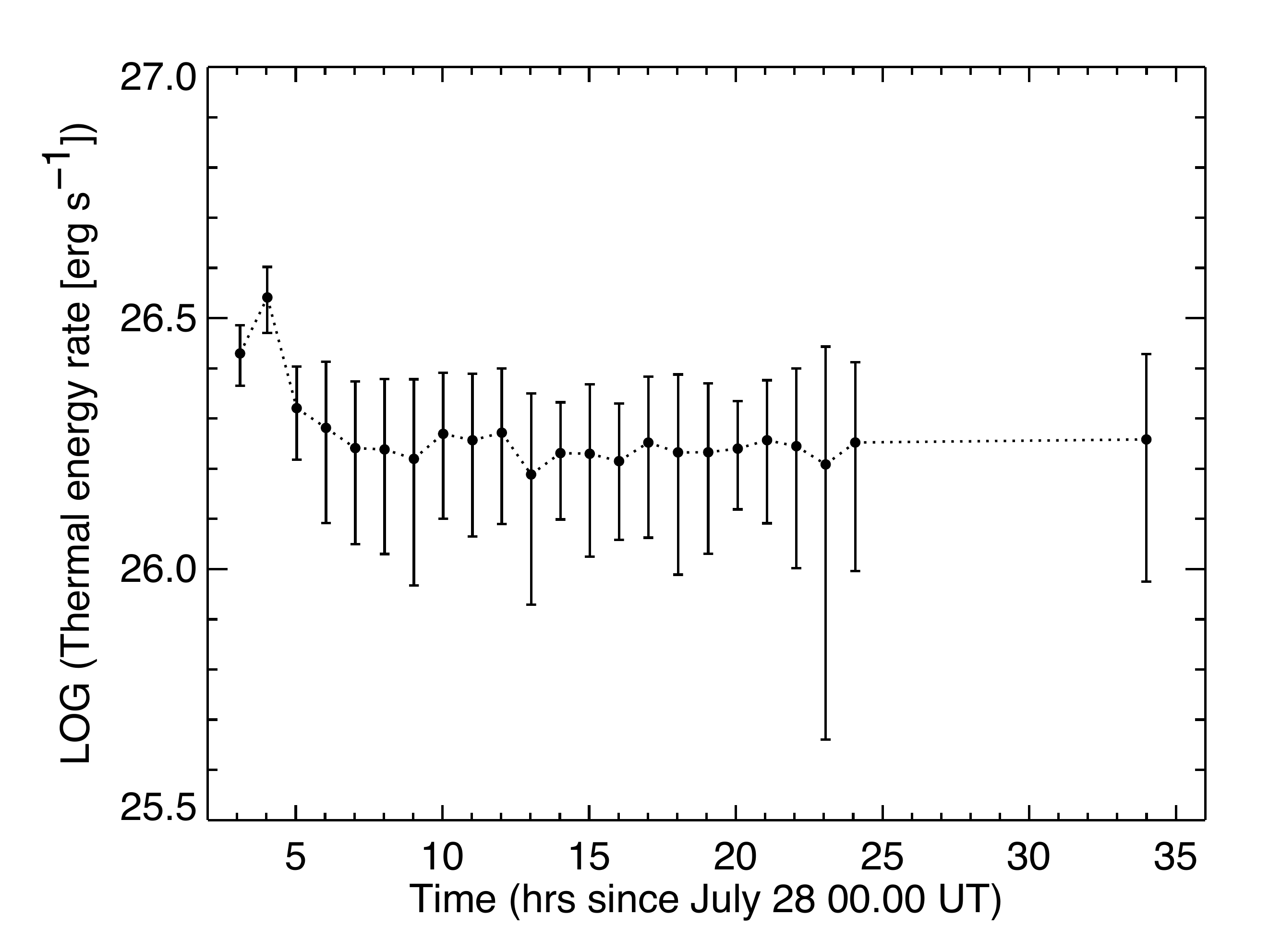}
	\caption{Plasma thermal energy rate along the CS at the UVCS altitude derived as described in the text, as function of the time.\label{uvcs_energy}}
\end{figure}

The evolution of the resulting thermal energy rate is plotted in Figure~\ref{uvcs_energy}.
The initial peak at 04:00~UT is correlated with the peak detected in \fexviii\ line emission.
After 06:00~UT, however, the thermal energy rate settles around the rather constant value, within the uncertainties, of $\sim 10^{26.3}$~erg~s$^{-1}$.
This implies a lower limit in the total energy needed to heat the current sheet of $\sim 2.2 \times 10^{31}$~erg and a upper limit of $4.2 \times 10^{31}$~erg, considering that the CS has been observed for at least a time of 30~hours (till 10:12~UT of July~29) and it disappeared in the next 29~hours (between 10:12~UT of July~29 and 15:10~UT of July~30).
The fact that the total energy flowing within the CS region is nearly constant for a relatively long time ($\sim 30$~hours) places important constraints on the mechanisms that can provide this energy in the lower corona at the base of the CS or in the intermediate corona sampled by UVCS (see later).

\subsection{Non-thermal velocities from spectral line profiles}

\begin{figure}[h]
\centering\includegraphics[width=\columnwidth]{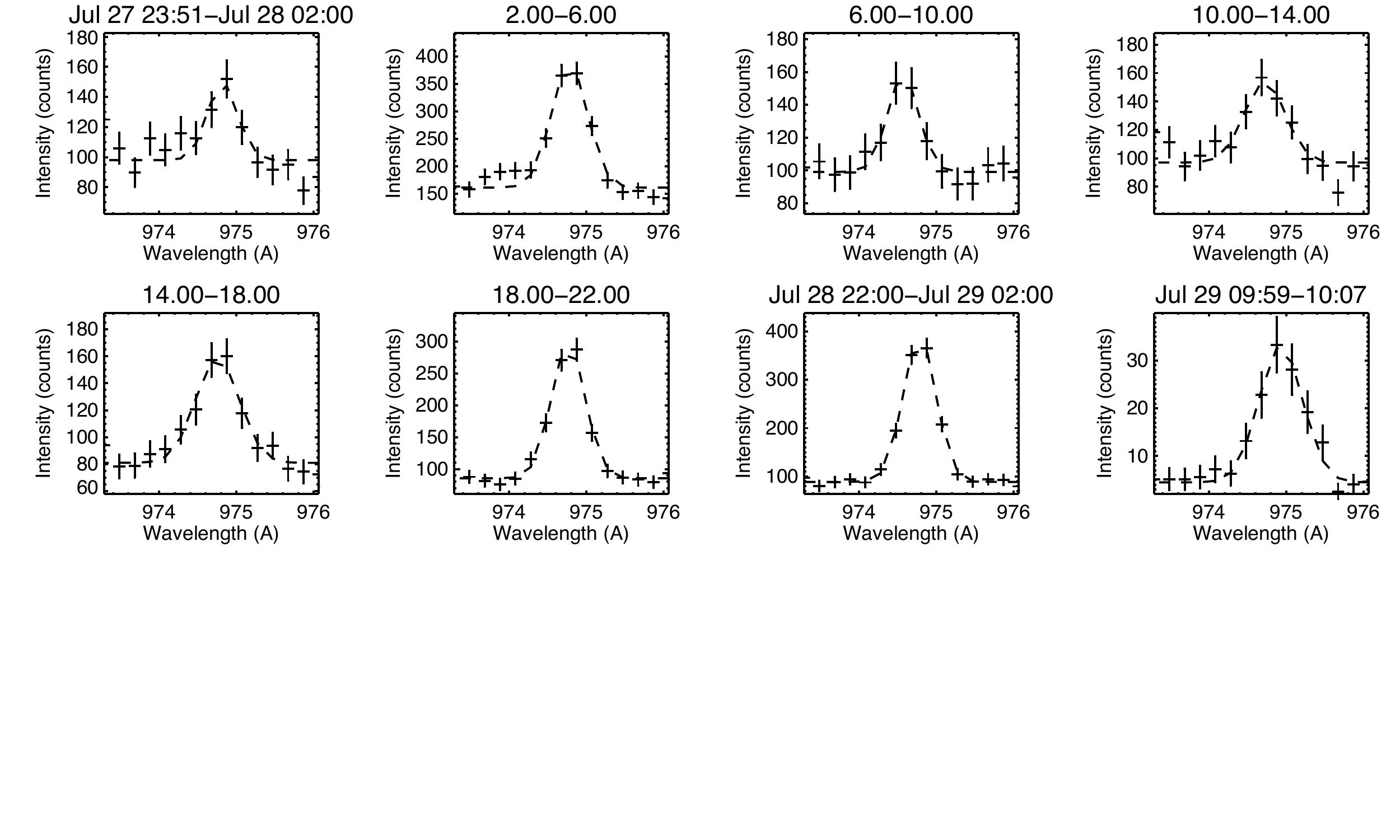}
\caption{Sequence of \fexviii\ line profiles averaged over 4~hours and over the spatial region corresponding to the CS. The first spectrum is obtained averaging only over 2~hours and 9~minutes (i.e. all the exposure available before the CME), while the last spectrum is obtained averaging only over 8~minutes (i.e. all the exposures available in the synoptic observation of July 29)\label{nth_spectra}}
\end{figure}

\begin{figure}[h]
\centering\includegraphics[width=0.75\columnwidth]{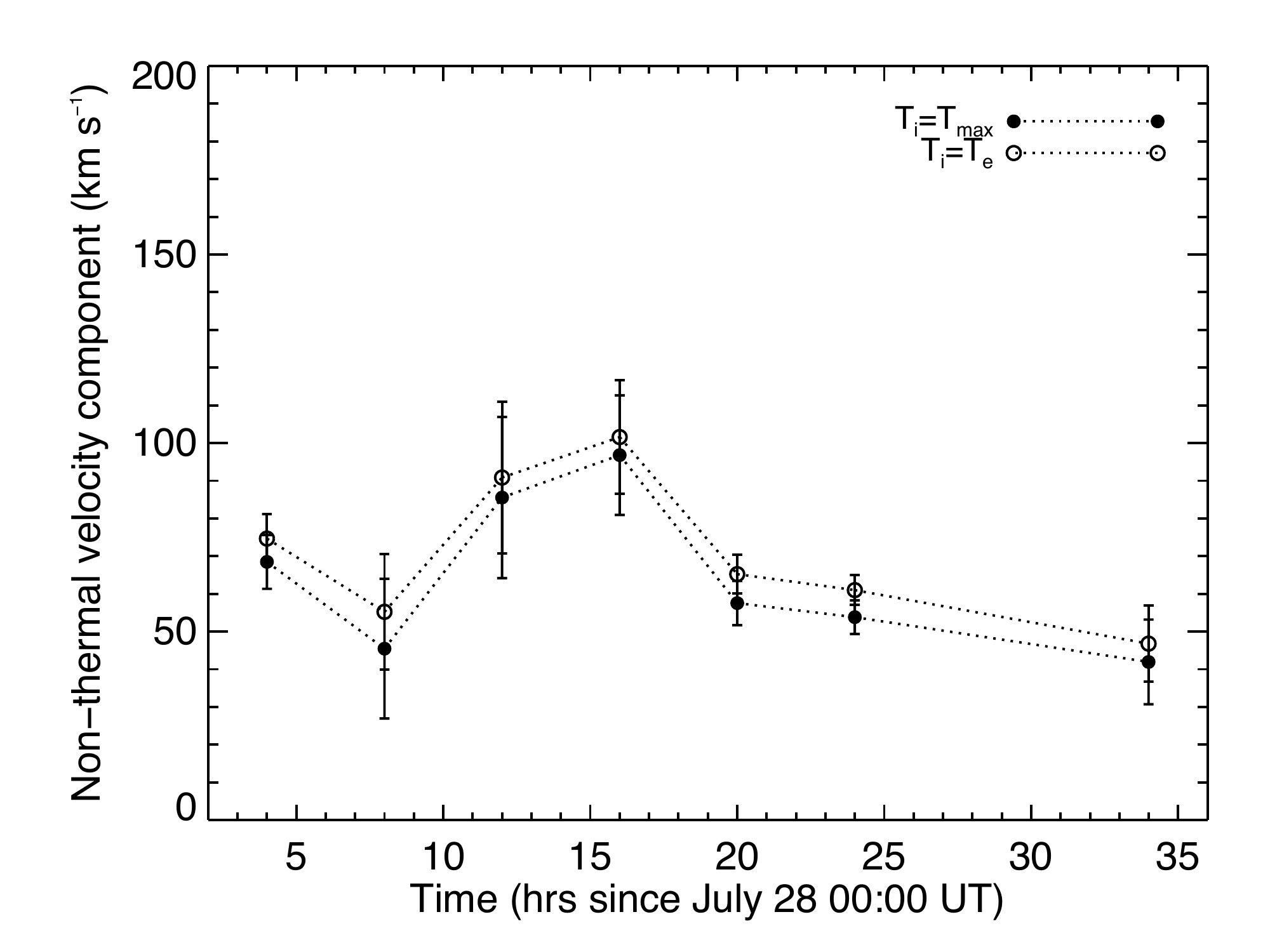}
\caption{Time evolution of the non-thermal velocity component of the plasma motions as derived from the spectral profile of the \fexviii\ line. Different curves correspond to different assumptions on the kinetic temperature of the emitting ions, as described in the text. \label{nth_speed}}
\end{figure}

A Gaussian fit to the \fexviii\ line profile is used to compute the FWHM of the line at different times during the observations.
The FWHM of a spectral line is related to the motions of the emitting ions.
If non-thermal motions are occurring with a RMS velocity $v_{\rm nth}$, then the effective FWHM, $\Delta\lambda_{\rm eff}$, is given by
\begin{equation}\label{deltaeff}
	\Delta\lambda_{\rm eff}=2\sqrt{\ln 2}\frac{\lambda_0}{c}\sqrt{\frac{2k_{\rm B}T_i}{m_i}+v_{\rm nth}^2},
\end{equation}
where $\lambda_0$ is the position of the centroid of the line, $T_i$ and $m_i$ the kinetic temperature and the mass of the emitting ion, respectively, $K_{\rm B}$ the Boltzmann constant, and $c$ the speed of the light.
To improve the statistics of the line profiles we combined several exposures and the resulting line profile are shown in Fig.~\ref{nth_spectra}. Then we corrected the observed FWHM of the line for the instrument profile width, $\sim 0.33$~\AA\ corresponding to $\sim 100$~km~s$^{-1}$, which takes into account the optical distortion, the slit width, and the spatial binning used in the observations \citep[we used the empirical formula given by][]{koh99}.
Non-thermal velocities have been finally computed from Eq.~(\ref{deltaeff}) in the two hypotheses of thermodynamic equilibrium, by assuming $T_i=T_e$ \citep[see][]{bem08}, and ionization equilibrium, by assuming $T_i=T_{\rm max}$, where $T_{\rm max} \simeq 10^{6.8}$~K is \fexviii\ peak temperature.
The resulting curves, which differ by less than 20\%, are plotted in Figure~\ref{nth_speed}.

The \fexviii\ line width exhibits a strong non-thermal component during the whole observation.
At the time the \fexviii\ emission peak (04:00~UT), $v_{\rm nth}$ is about 70~km~s$^{-1}$; later on, non-thermal velocities are always  around 60-70~km~s$^{-1}$ and slightly larger between 12:00~UT and 17:00~UT.
The values we found are generally higher than those obtained, for instance, by \citet{bem08}, but \citet{cia08} found even higher non-thermal speeds ($\sim 180$~km~s$^{-1}$) in their analysis of a post-CME current sheet.
These results suggest that turbulent broadening of the \fexviii\ line may be present in the CS.

\section{X-ray observations}

In order to understand the origin of the high-temperature plasma sampled by UVCS, we studied the evolution of the hard and soft X-ray sources observed during the same time interval at the base of the CS region in the lower corona.
It is usually believed that these emissions are due to Bremsstrahlung radiation emitted, respectively, by non-thermal electrons being accelerated in the reconnection region and electrons flowing along closed loops by chromospheric evaporation.
In agreement with this interpretation, the soft part of the X-ray spectra observed during solar flares is usually fitted with a thermal component at a fixed temperature ($\sim \mathrm{tens}$ of~MK), while the higher-energy part is fitted with a power law.
Hence, soft and hard X-ray emission are usually correlated, with the hard X-ray source (where reconnection releases energy) observed to lie above the soft X-ray loops (into which energy was released before). 
In this scenario, plasma being heated and accelerated in the reconnection point is also expected to propagate in the outer corona, possibly producing an unusually high-temperature emission.
Hence, what we want to check here for the first time is the possible agreement between energies required to explain the soft and hard X-ray sources observed at the base of the corona (by GOES and RHESSI) after a CME, and the energies required to explain the high-temperature emission observed at much higher altitudes (by SOHO/UVCS) during the same time interval.

\subsection{Soft X-ray data}\label{sxi_data}

Soft X-ray data analyzed here were acquired by the Solar X-ray Imager (SXI) onboard the GOES-12 satellite in a geo-synchronous orbit around the Earth. 
SXI provides full-disk, $512 \times 512$ images with a spatial resolution of approximately 10~arcsec FWHM, sampled with 5~arcsec pixels; images analyzed here were acquired alternatively with the ``Poly thin'', ``Be thin'', and ``Poly medium'' filters, sampling respectively the soft X-ray emission in the 6-65~\AA, 6-20~\AA\ and 6-60~\AA\ wavelength ranges, with an average image integration time of 3.0~s and a time cadence of one minute \citep[see][for a SXI instrument accurate description]{piz05,hil05}. 
During the time interval analyzed here (i.e. between July~28, 00:01:55~UT and July~29, 11:58:56~UT) SXI acquired 2095 exposures; only a very few of them (6) were acquired with the ``Be medium'' filter during the observation interval, hence these images were not analyzed. 
Images have been calibrated and coaligned with the standard routines provided within the \textit{SolarSoftware} package.

\begin{figure}[h]
	\centering\includegraphics[width=\textwidth]{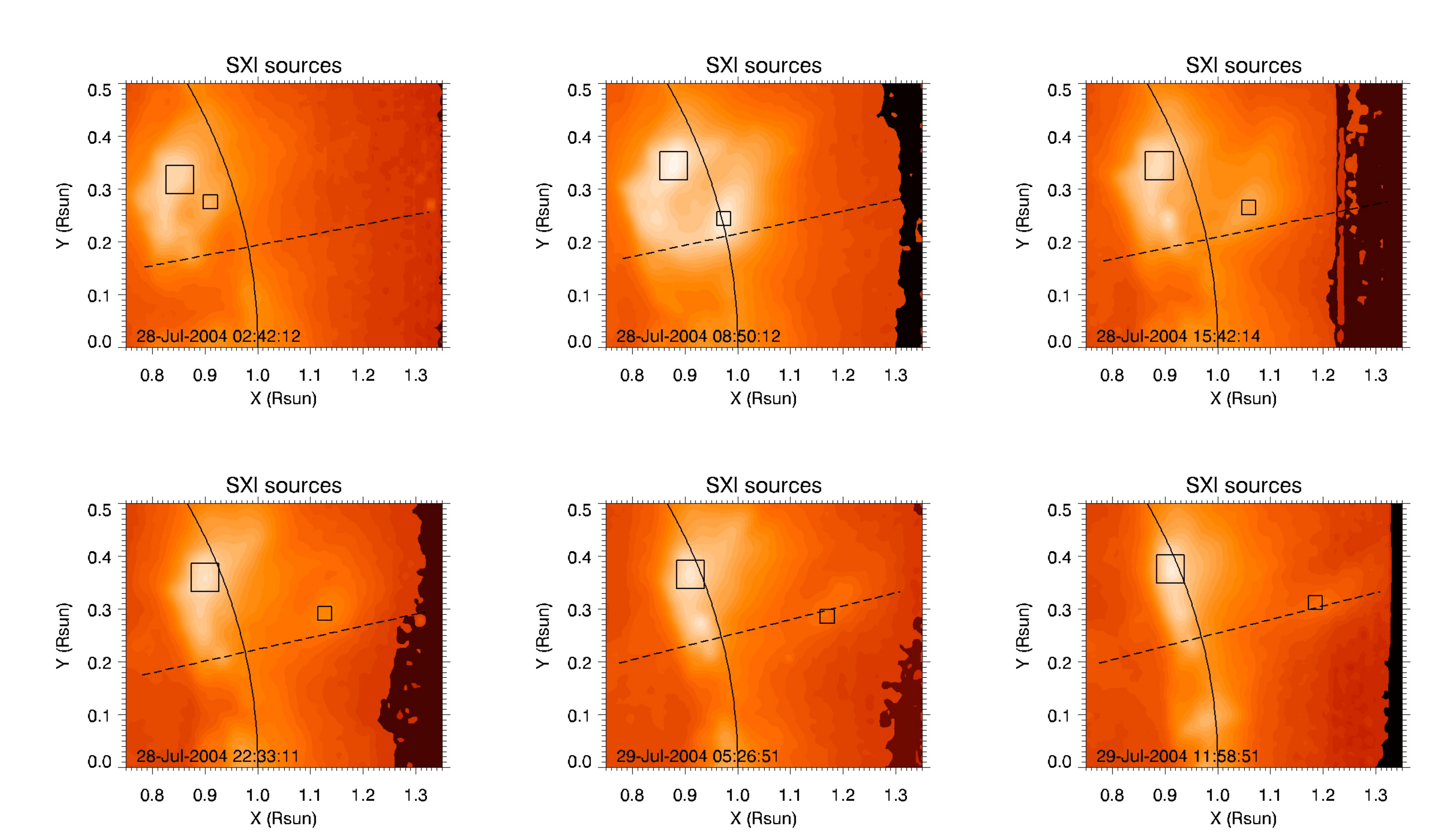}
	\caption{Sequence of GOES/SXI images acquired with the ``Poly thin'' filter showing the evolution of soft X-ray sources in the North-West quadrant from July~28, 02:42~UT to July~29, 11:58~UT. The boxes show the location of the disk and the rising sources relative to the latitude where the maximum temperature is observed during the same time interval in UVCS data at an altitude of 0.77~$R_\odot$ above the limb. In order to show the relative evolution of different soft X-ray emission sources, we used the same range of the color scale in all the plots.\label{sxi_sequence}}
\end{figure}

The sequence of SXI images acquired in the above time interval (Figure \ref{sxi_sequence}) clearly shows that on July~28, between 00:00~UT and $\sim 02$:00~UT, a single extended radiation source is located on disk, close to the North-West limb; no significant evolution is observed during the first 2~hours. 
Then, starting from $\sim$ 02:20~UT a secondary source appears, originally detaching from the extended source and then becoming an isolated emission source very close to the extended source since 02:42~UT. 
In order to better investigate the location and time evolution of the soft X-ray emission in these two sources, we automatically tracked their locations in the SXI images. 
This was done first by visually selecting their approximate location in the first frame we analyzed, and second by automatically selecting in the $(n+1)$-th frame the location of the emission's peak in an area centered on the coordinates found in the $n$-th frame; the areas we employed were $20 \times 20$ and $10 \times 10$~pixels large for the extended and compact sources, respectively. 
The soft X-ray emission observed with different filters has been then averaged frame by frame over these boxes in order to derive the light curves for the two sources.

\begin{figure}[h]
	\centering\includegraphics[width=\textwidth]{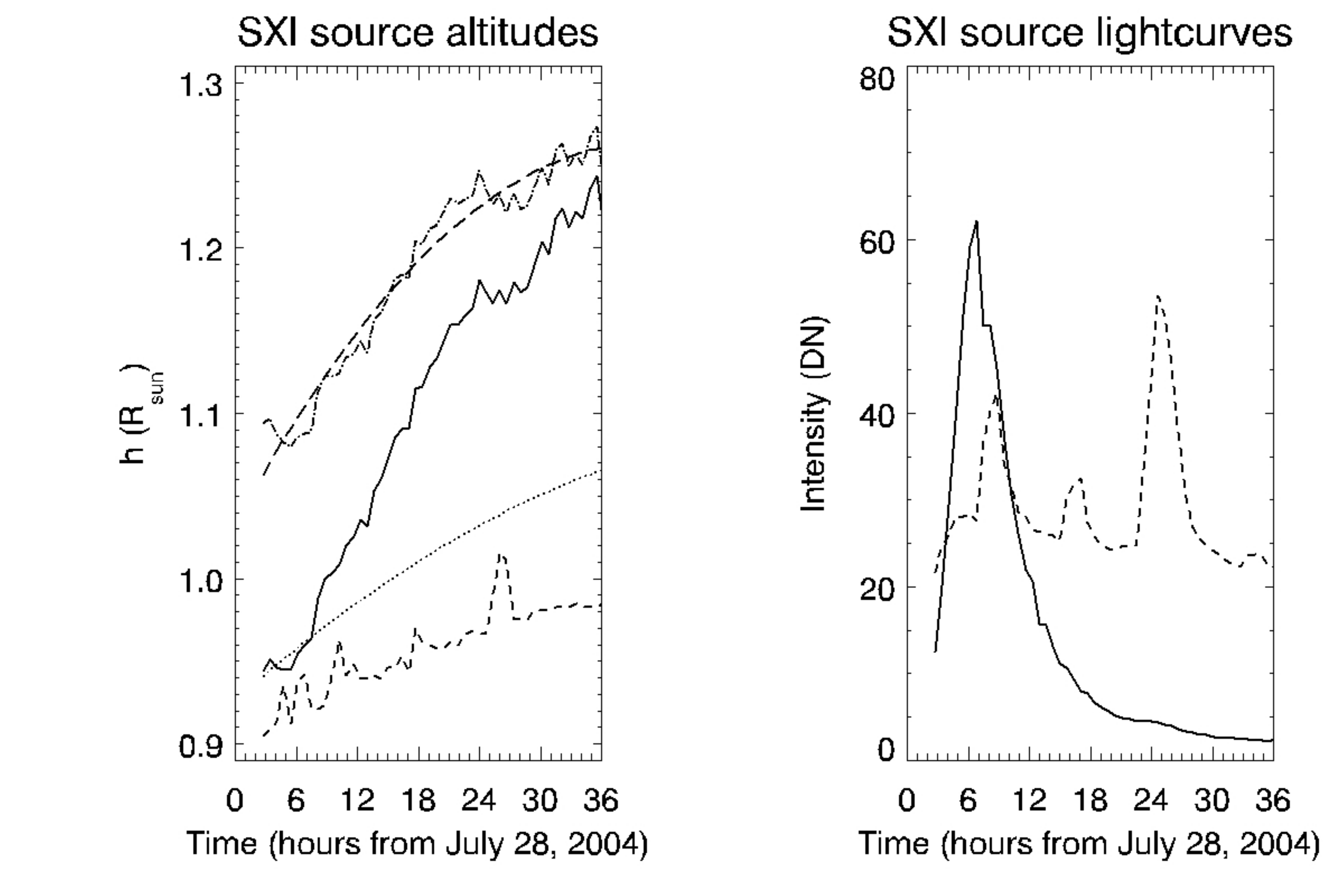}
	\caption{Left: measured altitudes (projected on the plane of the sky) of the disk (dashed line) and rising (solid line) soft X-ray sources. The dotted line shows the expected evolution of the projected altitude due only to solar rotation for a source lying at a real altitude of 0.1~$R_\odot$, while the dash-dotted line shows the rising source unprojected altitude reconstructed taking into account the solar rotation (see text). Right: measured soft X-ray light curves (observed with the ``Poly thin'' filter) for the disk (dashed) and the rising (solid line) sources.\label{sxi_curves}}
\end{figure}

Figure~\ref{sxi_curves} shows the resulting altitude vs. time curves (left panel) and the light curves (right panel) for the two soft X-ray sources. 
In particular, the left panel in the figure shows that the disk extended source appears to move simply by following the solar disk rotation and getting closer to the limb, while the second source appears to move faster outwards in time. 
A comparison between the projected altitudes observed for the rising source and those expected by solar rotation alone as a function of time (by assuming a source lying 0.1~$R_\odot$ above the solar surface) shows that the observed expansion is much larger than what implied by solar rotation alone. By simply assuming that, in first approximation, the source is expanding radially, it is possible to deproject the observed altitudes and to subtract from the observed expansion the motion due to the solar rotation, in order to derive the real expansion speed.

Results (dash-dotted line in the left panel of Figure~\ref{sxi_curves}) show that in the time interval between July~28, 02:42~UT and July~29, 12:00~UT the secondary source expanded from 1.08~$R_\odot$ to 1.26~$R_\odot$, with a very small average speed: second-order fit to the deprojected curve gives an initial velocity by 2.2~km~s$^{-1}$ around 02:42~UT, falling down to 0.3~km~s$^{-1}$ $\sim 33$~hours later; hence the rising motion is going stop $\sim 1.5$~days after the CME. During the same time interval, the projected latitude of the secondary source stays almost constant around $\simeq 13-14^\circ$ N, hence approximately $2-3^\circ$ northward of the projected latitude of the maximum plasma temperature seen by UVCS, moving closer to this latitude on July 29 (see Figure \ref{sxi_sequence}, bottom middle and right panels). 

An analysis of the light curves shows a very different behavior of the two sources: the soft X-ray emission from the disk extended source is almost constant, with a very few spikes likely due to continuous flaring activity of the source active region; on the contrary, the rising source shows a sharp increase over $\sim 4$~hours, reaching a peak in the emission around 06:50~UT, followed by a very slow decrease in the next $\sim 30$~hours. 
During the faster emission increase, the rising source moves mostly southwards and no significant radial expansion is observed, while, later on, the slower decay phase corresponds with the radial expansion phase of the source. 
The above considerations are valid for the average peak emission, but the light curves shown in Figure~\ref{sxi_curves} do not take into account that the disk source is compact at any time, while the rising source emission spreads over a larger area as the source moves outwards.

The rising soft X-rays source is located just above the EUV loops seen by EIT and shifted slightly northward: hence this source can easily be identified as the location of the newly reconnected loops, emitting at the beginning in hard X-rays and later on in EUV. The quite good coalignement between this source and the location of the temperature peack seen by UVCS indicates that the CS region extends almost radially above the post-CME loops.

\subsection{RHESSI hard X-ray observations}

\begin{figure}[h]
\centering
\includegraphics[width=\textwidth]{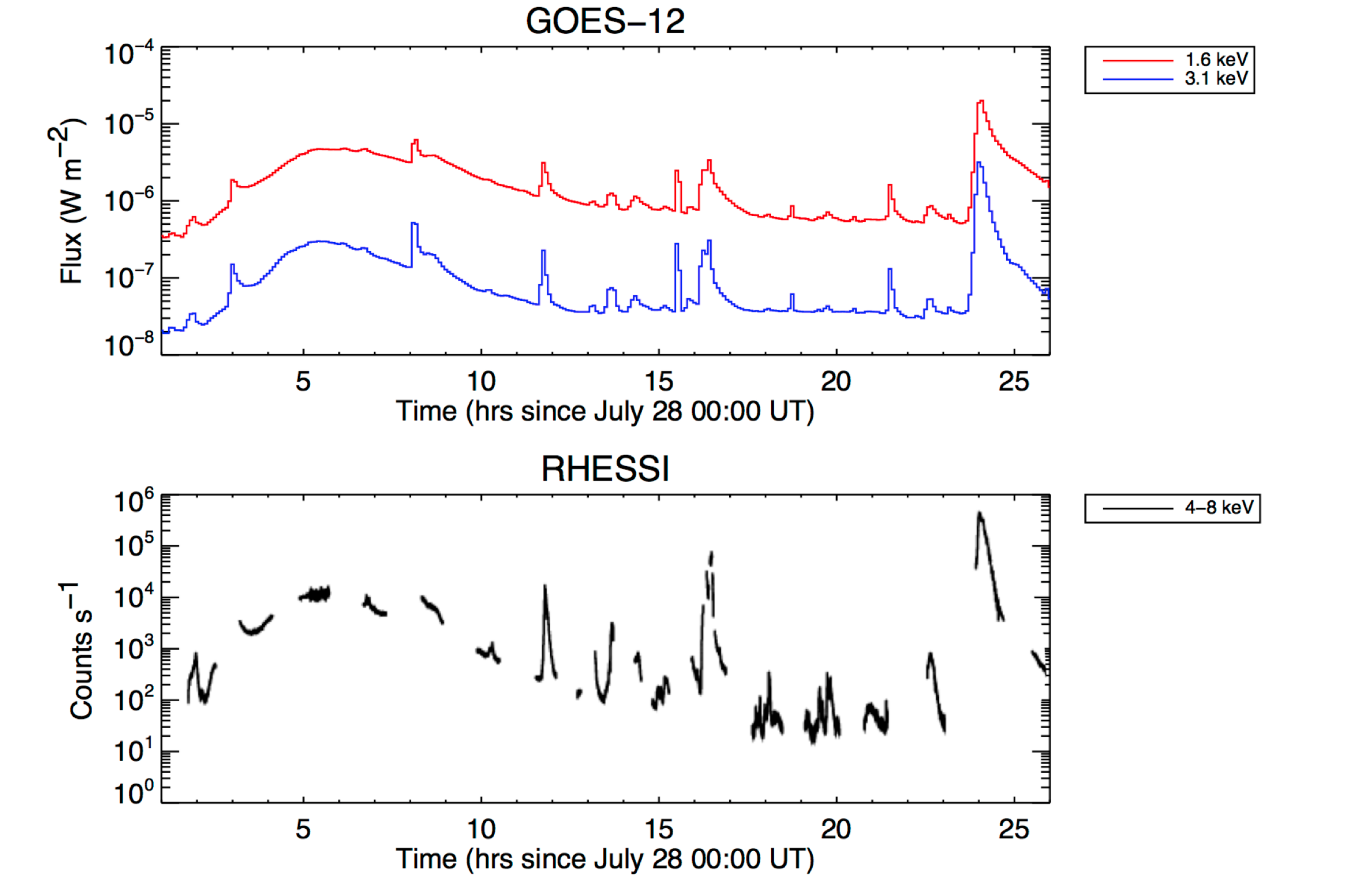}
\caption{GOES-12 fluxes in the 1.6~keV and 3.1~keV bands (top panel) and RHESSI count rates in the 4-8~keV energy range (bottom panel).\label{goes_rhessi}}
\end{figure}

To study the hottest plasma and potentially non-thermal emissions associated with the current sheet discussed in this paper, observations taken by the Reuven Ramaty High Energy Solar Spectroscopic Imager \citep[RHESSI,][]{lin02} in the $>3$~keV hard X-ray have been analyzed. 
The HXR event associated with the current sheet is very similar as the event discussed by \citet{sai09}.
The RHESSI spacecraft has a day-night cycle and therefore only provides partial coverage of the long-duration event of 2004 July~28.  
Nevertheless, significant parts of all phases associated with the hot current sheet are captured by RHESSI starting with the onset of the HXR emission, parts of the impulsive phase, and several episodes during the decay (see Figure~\ref{goes_rhessi}). 
During this long-duration event, several additional, most likely independent smaller scale flares were also seen by RHESSI, all coming from the same location as the SXR events discussed in Section~\ref{sxi_data}. 
Besides the unusually long duration, the most striking feature of the HXR observations is the absence of non-thermal HXR emissions. 
The RHESSI spectrum is thermal, without any signatures of a flattening at high energies as normally seen during the impulsive phase of flares \citep[e.g.][]{ray12}, at least during the times when RHESSI is observing the Sun. 
Hence, the magnetic reconnection associated with the hot current sheet seems to be very inefficient in accelerating electrons, contrary to typical flares. 

\begin{figure}[h]
\centering
\includegraphics[width=\textwidth]{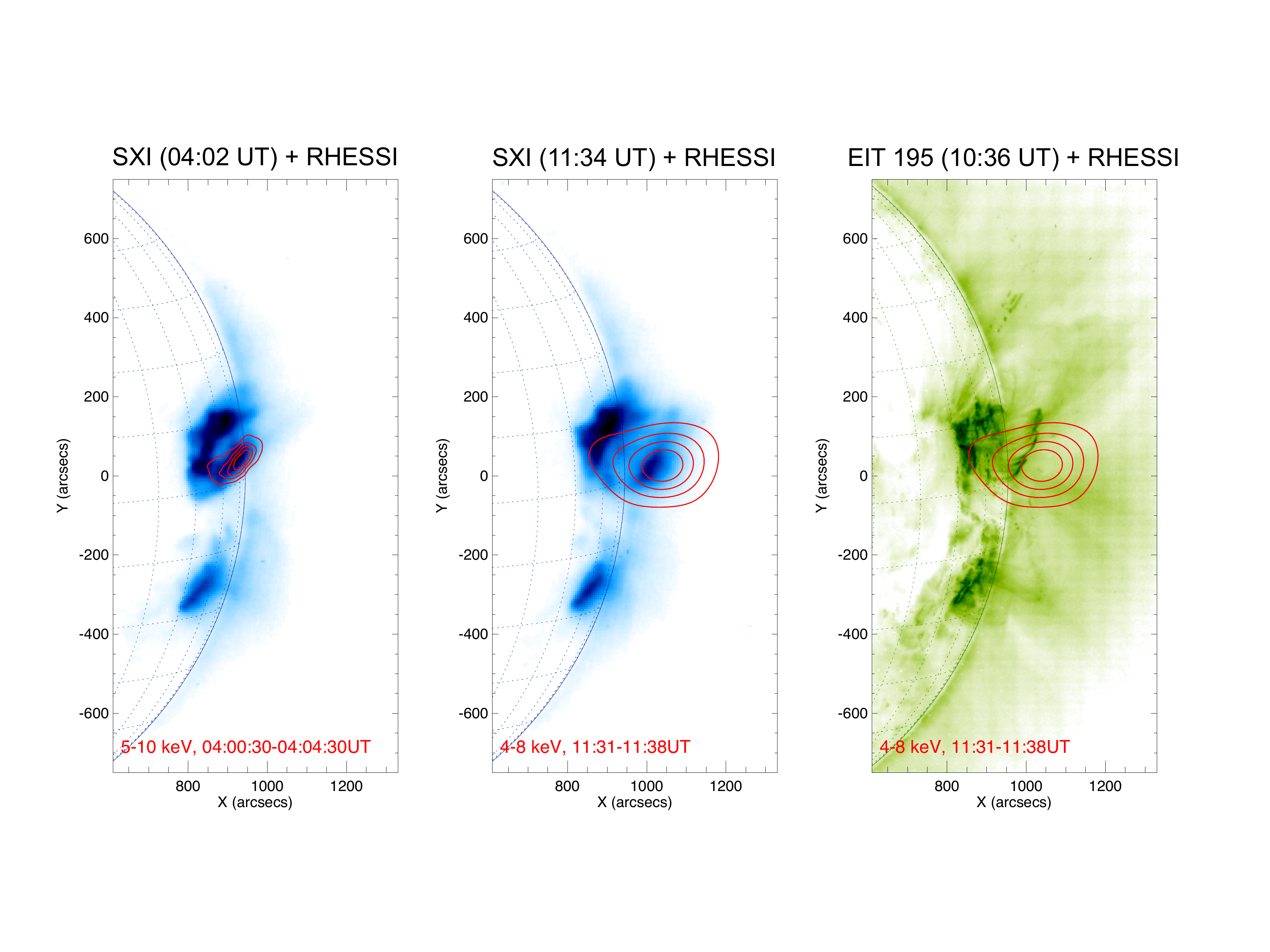}
\caption{Left and middle panels: GOES/SXI images of the source soft X-ray emission region acquired at 04:02~UT and 11:34~UT on 2004 July~28 with the ``Poly thin'' filter, with RHESSI hard X-ray emission contours at the same times (4~minutes and 7~minutes exposures, respectively; red circles). Right panel: SOHO/EIT 195~\AA\ image taken at 10:36~UT on the same day, with RHESSI 4-8~keV contour (4 minutes exposures; red circles).\label{rhessi_composite}}
\end{figure}

We used the RHESSI CLEAN algorithm \citep{hur02} to make hard X-ray images in the 5-10~keV that correspond to the rising source seen in SXR. 
The HXR images outline the top of post-CME loops and show the same rising motion as seen in SXR (Figure~\ref{rhessi_composite}). 
The location of the HXR source is above the EUV and SXR loops, indicating the typical structure of hotter loops above cooler loops \citep[e.g.][]{gal02}, as expected from the simplest reconnection models.  
Compared to general HXR flare loops, the source is unusually large with $30''\times 90''$ at the peak of the \fexviii\ emission, and it expands in time to reach a source size of up to $\sim 180''$, the upper limit what RHESSI can image \citep[for details on RHESSI source size determination see][]{den09}.
RHESSI spectral fitting \citep[e.g.][]{smi02} reveals a temperature of $\sim 18$~MK at the peak of the \fexviii\ emission ($\sim 04$:00~UT) with an emission measure $EM=n^2V$ around $5 \times 10^{46}$~cm$^{-3}$. 
Hence, the post-CME loops are significantly hotter than the current sheet at higher altitudes. 
Assuming the thickness of the HXR source to be equal to the width, the volume is $V=30'' \times 30'' \times 90''$, and the density of the HXR loops becomes $n \approx 1 \times 10^{-9}$~cm$^{-3}$ at the \fexviii\ peak.
This is an unusually low density for a flare loop \citep[e.g.][]{cas13}.
The total energy content in the loop of $3 kT n V$ \citep[e.g.][]{han08} can then be estimated to $3 \times 10^{29}$~erg.
This energy has to be understood as the instantaneous energy content within the thermal plasma. 
Hence, it is a lower limit to the energy that is need to heat the plasma in the loop to the current state; the actual energy that is need is higher by the amount of the total losses (mainly by thermal conduction) that occur during the heating process. 
The time evolution of the spectral HXR observations shows that the peak temperature is reached around $\sim 05$:00~UT with $\sim 22$~MK ($EM \approx 9 \times 10^{46}$~cm$^{-3}$), and the loops cool down afterwards to around $\sim 9$~MK ($EM \approx 5 \times 10^{47}$~cm$^{-3}$) around 11:00~UT. 

\begin{deluxetable}{p{2.8cm}p{5.3cm}p{2.5cm}p{2.5cm}p{5.8cm}}
\tabletypesize{\scriptsize}
\tablewidth{21cm}
\tablecolumns{5}
\tablecaption{Overall description of the observed features and proposed interpretation.\label{table_events}}
\tablehead{
\colhead{\textbf{Time}} & \colhead{\textbf{Event}} & \colhead{\textbf{Instrument}} & \colhead{\textbf{Wavelength}} & \colhead{\textbf{Interpretation}}}
\rotate
\startdata
\mbox{July~28, 2004} \mbox{$\sim 02$:30~UT} & Beginning of gradual rise in the GOES X-ray emission & GOES & X-rays & \\[0.5cm]
\mbox{$\sim 02$:30~UT} & Approximate CME start time back-extrapolated from $h$ vs. $t$ curves & SOHO/LASCO-C2 & White light & \\[0.5cm]
\mbox{$\sim 02$:42~UT} & New soft X-ray source starts to brighten southern of the AR & GOES12/SXI & Soft X-rays images (6-65~\AA) & Emission from footpoints of newly reconnected loops, classical signature of post-CME reconnection (bremsstrahlung X-ray emission from accelerated electrons) \\[0.5cm]
\mbox{03:14~UT} & New hard X-ray source starts to brighten southern of the AR & RHESSI & Hard X-rays images (5-10~keV) & Emission from foot-points of newly reconnected loops, classical signature of post-CME reconnection (bremsstrahlung X-ray emission from accelerated electrons) \\[0.5cm]
\mbox{03:30~UT} & Coronal white light emission starts to rise at 2~$R_\odot$ at the latitude of the CME & SOHO/LASCO-C2 & White light & Compression of the overlying corona due to the expansion of CME plasmoid (Thompson scattering from enhanced density electrons) \\[0.5cm]
\tablebreak
\mbox{04:00~UT} & Thermal Hard X-ray spectrum without non-thermal emission & RHESSI & Hard X-rays spectrum (5-100~keV) & Classical chromospheric evaporation in the post-CME reconnected loops; no significant particle acceleration \\[0.5cm]
\mbox{03:30-04:30~UT} & Peak in the high temperature UV coronal emission at 1.77~$R_\odot$ & SOHO/UVCS & UV (974~\AA) & Transit of the CME core in the UVCS FOV (UV emission of high-T plasma heated by reconnection and embedded in the plasmoid) \\[0.5cm]
\mbox{$\sim 04$:00-04:30~UT} & Back extrapolated transit time of CME core in the UVCS FOV at 1.77~$R_\odot$ & SOHO/LASCO-C2 & White light & See above \\[0.5cm]
\mbox{$>04$:36~UT} & EUV Post flare loops rising from the southward end of the AR & SOHO/EIT & EUV (195~\AA) & Emission from newly reconnected loops, signature of post-CME reconnection \\[0.5cm]
\tablebreak
\mbox{05:06~UT} \mbox{(05:30~UT)} & CME core visible in LASCO-C2 around 3.48~$R_\odot$ (4.55~$R_\odot$) & SOHO/LASCO-C2 & White light & CME propagation phase \\[0.5cm]
\mbox{06:50~UT} & Peak in the soft X-rays from the new source and beginning of its outward propagation & GOES12/SXI & Soft X-rays images (6-65~\AA) & Emission from the top of newly reconnected loops, signature of post-CME reconnection (bremsstrahlung X-ray emission from accelerated electrons) \\[0.5cm]
\mbox{$>07$:00~UT} & Beginning of the outward radial propagation of new hard X-ray source & RHESSI & Hard X-rays images (5-10~keV) & Post-CME reconnection proceeds at higher altitudes \\[0.5cm]
\mbox{$>07$:00~UT} & Progressive decay of secondary source soft X-rays emission and outward radial propagation (1.10-1.25~$R_\odot$) & GOES12/SXI & Soft X-rays images (6-65~\AA) & See above \\[0.5cm]
\mbox{$>07$:00~UT} & Rising UV high-temperature coronal emission at 1.77~$R_\odot$, lasting more than 1.5~days & SOHO/UVCS & UV (974~\AA) & Post-CME emission from plasma heated by turbulent reconnection in the current sheet \\[0.5cm]
\mbox{July~29} \mbox{$\sim 10$:00~UT} & High temperature UV coronal emission still present at 1.77~$R_\odot$, still rising in time $\sim 30$~hours after the CME & SOHO/UVCS & UV (974~\AA) & Post-CME turbulent reconnection in the current sheet still efficiently working \\[0.5cm]
\mbox{July~29} \mbox{$\sim 11$:00-12:00~UT} & End of the outward propagation of the soft X-ray rising source and disappearance & GOES12/SXI & Soft X-rays & Decay of post-CME reconnection at the base of the current sheet \\[0.5cm]
\mbox{July~30} \mbox{$\sim 15$:25~UT} & Absence of high temperature UV coronal emission at 1.77~$R_\odot$ & SOHO/UVCS & UV (974~\AA) & Decay of post-CME turbulent reconnection in the current sheet in the previous $\sim 30$~hours \\
\enddata
\end{deluxetable}

\section{Discussion and conclusions}

The variety of emission features associated with the CME erupted on July~28, 2004, and detected at different times by SOHO/UVCS and LASCO, RHESSI, and GOES~12/SXI are summarized in Table~\ref{table_events}.

The approximate CME starting time, extrapolated from altitude-vs.-time diagrams of white-light features identified in LASCO~C2 images, is $\sim 02$:30~UT.
The soft and hard X-ray sources appearing later on (after 02:42~UT) on the solar disk can be then interpreted as the emission coming from the newly reconnected post-CME loops and produced by the electron accelerated in the process.
This is confirmed by the good agreement found between the location of the X-ray sources and the position of the footpoints of the EUV loops observed at the same time in SOHO/EIT 195~\AA\ images.
Note that, due to the high inclination of the post-CME loops with respect to the LOS, EIT images show only one single, extended spot of bright emission, and not two separated spots as expected for the two footpoints.

After 03:30~UT, the CME front and then the core entered the LASCO~C2 field of view.
The white-light emission is produced by Thompson scattering from the enhanced electron density of the plasma, produced in turn by the compression of the coronal plasma by the rapid expansion of the CME.
Between 03:30~UT and 04:30~UT, UVCS observed at 1.77~$R_\odot$ a short, pronounced peak in the \fexviii\ $\lambda$974~\AA\ spectral line, corresponding to high temperature ($\sim6.5$~MK) plasma emission.
Using SOHO/LASCO~C2 images, we extrapolated back the transit time of the CME core at the altitude of the UVCS FOV and found that this time ($\sim04$:00~UT) is in very good agreement with the time of the \fexviii\ emission peak.
Hence, the well-defined peak observed in the \fexviii\ spectral line by UVCS can be interpreted as the UV emission of the high-temperature plasma heated by reconnection and embedded in the CME core plasmoid.
At the same time, RHESSI detected hard X-ray thermal emission, without a non-thermal counterpart, located near the top of the post-CME EUV loops, which was probably produced by chromospheric evaporation of the footpoint plasma heated by the particles accelerated during the reconnection of the loop system.

Later on, both RHESSI and GOES/SXI detected hard and soft X-ray emission from the top of the post-CME loop system.
This emission is observed to move progressively at higher altitudes and is probably the signature of the magnetic reconnection taking place above the top of the loops \citep[e.g.,][]{sai09}.
Simultaneously, starting at 07:00~UT, UVCS observed a progressive rise in UV high-temperature coronal emission in the \fexviii\ and \sixii\ lines.
The overall characteristics of the emission seen by UVCS at this stage, i.e., (1) a very long duration ($> 1.5$~days), (2) a relatively narrow spatial distribution along the UVCS slit ($\sim 10^\circ$), and (3) the presence of significant non-thermal broadening of the lines (corresponding to non-thermal velocities up to $\sim 100$~km~s$^{-1}$), are in good agreement with those previously derived from UVCS observations of post-CME current sheets \citep[see, e.g.,][]{bem06,cia08}.

In particular, the temporal evolution of the electron temperature and density, derived form the line intensity ratio of the \fexviii\ and \sixii\ spectral lines, shows that, at the altitude of the UVCS slit, the CS plasma has been heated continuously for at least 30~hours at a relatively constant energy rate.
The origin of this long-term energy rate flowing through the CS is unknown, but two explanations are possible: (1) CS thermal energy is produced by dissipation of magnetic energy through Petschek-like reconnection occurring mainly at the base of the CS, i.e., at the location of the HXR source above the top of the post-CME loops, and then it is transferred via thermal conduction and/or plasma motions at higher altitudes along the CS \citep[e.g.][]{ko10,vrs09}, or (2) CS thermal energy originates from multiple reconnections occurring at small scale levels in a turbulent ambient plasma located in the intermediate corona around the altitude sampled by UVCS \citep[][]{mck13,bem08}.

Our results show that, during the entire event, the instantaneous energy content of the HXR source increases constantly, reaching $\sim 2 \times 10^{30}$ erg around 11:00~UT. 
Hence, the HXR source is being constantly heated at a level larger than the combined conductive plus radiative losses. 
The losses are likely dominated by conductive losses through the chromosphere (see also Saint-Hilaire et al. 2009), while radiative losses can be neglected. 
The conductive loss time scale for the HXR source are of the order of $\tau_{\rm C}\approx10^3$s \citep[for details see][]{sai09}, so the total losses can be estimated as $E \Delta t /\tau_c $, where $E$ is the averaged thermal energy content and $\Delta t$ is the total duration. 
Using $E \simeq 10^{30}$ erg, $\tau_{\rm C} \simeq 10^3$ s, and $\Delta t \simeq 7$ hours, the total released energy becomes of the order of $10^{32}$ erg and is actually comparable to the energy released in a large X-class flare \citep[e.g.][]{ems12}.
The energy release in this event seems to be less impulsive than in regular flares without electron acceleration, but long-lasting at a constantly high level.

Hence the amount of energy embedded in the HXR source is larger by a factor $\sim 2-5$ than the total thermal energy of current sheet as estimated from UVCS data ($\sim 2.2-4.2 \times 10^{31}$~erg). Probably this last quantity is slightly underestimated, because 1) of the small value we assumed for the outflow velocity along the CS ($v_{\rm out} \approx 54$~km~s$^{-1}$), and 2) because of possible effects related to the assumption of ionization equilibrium (see below). In any case, plasma embedded in the HXR source could in principle provide sufficient energy to heat the CS plasma sampled by UVCS. Nevertheless, energy in the HXR source is likely to be trapped in the newly reconnected post-CME loops and to be more efficiently transferred downward, since energy transport via thermal conduction is strongly inefficient towards upper levels.
For instance, \citet{ree10} investigated coronal energy flow during a simulated CME and associated current sheet, with a detailed energy diagnostics.
They showed that toward the end of the eruption, i.e., during the formation and evolution of the CS, the majority of the energy flux along the CS in the upper direction is due to kinetic energy flux. This elevated kinetic energy flux is attributable to the reconnection outflow jet. Thermal conduction along the current sheet is strongly inhibited toward the upper direction, while it remains the dominant form of energy transport toward the lower direction. The energy flow through the lower tip of the CS is available for the heating of the post-flare loops and is responsible for the X-ray emission as demonstrated by \citet{ree10}. Hence, energy we may argue that only a small fraction of thermal energy produced at the location of the HXR source will reach the altitude sampled by UVCS (0.77 R$_\odot$ above the limb). In addition, the fact that the SXR source fades and its rising motion is going to stop after $\sim 1.5$~days after the CME, while the UVCS high-temperature emission is still increasing with time more than 30 hours after the event, suggests that the origin for the high-temperature emission observed by UVCS is not directly related with the energy released in the post-CME reconnection occurring above the top of the magnetic arcades.

The alternative heating process we suggest here is turbulent reconnection occurring in the elongated current sheet above the post-CME loops. \citet{ni12} have recently demonstrated that, when temperature-dependent diffusivity and anisotropic thermal conduction is assumed in MHD simulations of reconnecting current sheets, one of the effects is the appearance of secondary magnetic islands, or plasmoids, along the CS, where the heat is transferred and where it gets trapped. If plasma turbulence turns out to be efficient inside the current sheet, as confirmed by the detection of significant non-thermal motions derived from the \fexviii\ spectral line width, many of these magnetic islands are likely to form along the CS due to the plasma micro-instabilities.
We can argue that, in this way, a large fraction of the thermal energy produced through magnetic reconnections occurring inside the many ``microscopic'' reconnection sites could be trapped in the CS and provide the heating to maintain the high temperatures detected by UVCS. This hypothesis is strengthened by the recent work of \citet{mck13}, who analyzed high-resolution observations of the solar corona performed with Hinode/XRT and SDO/AIA in the region associated with two post-CME current sheets and found evidence of significant shears in velocity, giving the appearance of vortices and stagnations. We notice that resulting non-thermal velocities correspond to a non-thermal energy density $E_{\rm nth} \sim 2 \times 10^{-3} \rm erg cm^{-3}$ which is much smaller than the average thermal energy of the CS plasma $E_{\rm th} \sim 8 \times 10^{-2} \rm erg cm^{-3}$.

Hence, these findings demonstrate that (1) turbulent effects are present in or very near the CSs in CME-associated eruptive flares and (2) turbulence very likely triggers modifications on the magnetic field, and possibly on the current sheets, including contortions to the magnetic field and cascades to small length scales \citep[see also][]{bem08}. Notice that the development of plasma turbulence inside the CS makes also not possible at the present state of knowledge to estimate the conductive losses along the CS, as we did for the HXR source. In fact, simulations by Ni et al. 2012 (ApJ, 758, 20) focused on the impact of thermal conduction in the evolution of CS during the development of plasmoid instabilities: the authors found that the thermal conduction becomes anisotropic, leading to a transfer of thermal energy from the reconnection points into the plasmoids, where the thermal energy is trapped. Then, it is still an open issue to understand how much of the energy being trapped in the plasmoids is eventually transferred along the CS by the thermal conduction in the direction of the vertical guiding magnetic field.

Before concluding, we remind that throughout this work we assumed ionization equilibrium in the computation of plasma parameters from the observed \fexviii\ and \sixii\ line intensities. In several conditions (e.g., in low-density, high-temperature plasmas characterized by significant bulk motions) this may be an unrealistic assumption. In particular, in observed current sheets, the ionization time scale of the ions may be longer than the dynamical time scale associated with the high-speed reconnection outflow. Recently, \citet{she13} analyzed the time-dependent ionization inside a large-scale and dynamically developed CME current sheet using the CME eruption model of \citet{ree10} and computed the intensities of \fexviii\ 975~\AA, \sixii\ 499~\AA, and other UV lines, as expected from UVCS observations made at a heliocentric distance of 1.77~$R_\odot$ in the direction across the CS. Their results show that the difference between equilibrium ionization and non-equilibrium ionization causes temperatures to be underestimated at those heights by about a factor of 2 if equilibrium is assumed in interpreting the data. This indicates that our estimates of the electron temperature inside the CS and, in particular, the amount of thermal energy flowing along the CS may be underestimated by a factor $\sim 2$: in any case this small variation does not affect the main results of this work. In addition, \citet{she13} found that inside the current sheet, the \fexviii\ profile exhibits a maximum peak located in the center of the CS, while the profiles of the \sixii\ and of the other UV lines they considered are less bright at the current sheet center then in the ambient corona.
Our spatial distributions of line intensities across the CS direction agree well with these results.

\acknowledgments
\begin{acknowledgements}
This work has received funding from the European Commission’s Seventh Framework Programme (FP7/2007-2013) under the grant agreement SWIFF (project no. 263340, {\tt www.swiff.eu}). 
It was also supported by the Swiss Science Foundation (200021-140308), by the European Commission through HESPE (FP7-SPACE-2010-263086), and through NASA contract NAS 5-98033 for RHESSI.
\end{acknowledgements}

\newpage

\newpage

\newpage

\newpage

\newpage

\newpage

\newpage

\newpage

\newpage

\newpage

\newpage

\newpage

\newpage

\newpage

\newpage

\end{document}